\documentclass[paper,11pt]{JHEP}
\pdfoutput=1
\usepackage[centertags]{amsmath}
\usepackage{amsfonts} \usepackage{amssymb} \usepackage{amsthm}
\allowdisplaybreaks[1]
\usepackage{graphicx}

\newcommand{\ii}{\mathrm{i}}

\newcommand{\cC}{{\mathcal{C}}}

\newcommand{\one}{{\rm 1\kern -.9mm l}}

\newcommand{\ft}[2]{{\textstyle\frac{#1}{#2}}}

\newcommand{\rme}{\mathrm{e}}

\newdimen\tableauside\tableauside=1.0ex
\newdimen\tableaurule\tableaurule=0.4pt
\newdimen\tableaustep
\def\phantomhrule#1{\hbox{\vbox to0pt{\hrule height\tableaurule
width#1\vss}}}
\def\phantomvrule#1{\vbox{\hbox to0pt{\vrule width\tableaurule
height#1\hss}}}
\def\sqr{\vbox{%
  \phantomhrule\tableaustep
\hbox{\phantomvrule\tableaustep\kern\tableaustep\phantomvrule\tableaustep}%
  \hbox{\vbox{\phantomhrule\tableauside}\kern-\tableaurule}}}
\def\squares#1{\hbox{\count0=#1\noindent\loop\sqr
  \advance\count0 by-1 \ifnum\count0>0\repeat}}
\def\tableau#1{\vcenter{\offinterlineskip
  \tableaustep=\tableauside\advance\tableaustep by-\tableaurule
  \kern\normallineskip\hbox
    {\kern\normallineskip\vbox
      {\gettableau#1 0 }%
     \kern\normallineskip\kern\tableaurule}%
  \kern\normallineskip\kern\tableaurule}}
\def\gettableau#1 {\ifnum#1=0\let\next=\null\else
  \squares{#1}\let\next=\gettableau\fi\next}
\def\XXint#1#2#3{{\setbox0=\hbox{$#1{#2#3}{\int}$}
     \vcenter{\hbox{$#2#3$}}\kern-.5\wd0}}

\title{Modular anomaly equations and S-duality in ${\mathcal N=2}$ conformal SQCD
}
\author{S. K. Ashok$^{1}$, M. Bill\'o$^{2}$, E. Dell'Aquila$^{1}$, M. Frau$^{2}$, A. Lerda$^{3,2}$, M. Raman$^{1}$
\\
\vskip 0.2cm
$^1$ Institute of Mathematical Sciences \\
   C. I. T. Campus, Taramani\\
   Chennai, India 600113\\
\vskip 0.2cm
$^2$ Universit\`a di Torino, Dipartimento di Fisica
\\ and I.N.F.N. - sezione di Torino, 
Via P. Giuria 1, I-10125 Torino, Italy\\
\vskip 0.2cm
$^3$Universit\`a del Piemonte Orientale, Dipartimento di Scienze e Innovazione Tecnologica, \\
and I.N.F.N. - Gruppo Collegato di Alessandria - sezione di Torino\\
Viale T. Michel  11, I-15121 Alessandria, Italy\\
\vspace{0.35cm}
\email{sashok,edellaquila,madhur@imsc.res.in; billo,frau,lerda@to.infn.it} 
}
\abstract{We use localization techniques to study the non-perturbative properties of an $\mathcal{N}=2$ superconformal gauge theory with gauge group SU(3) and six fundamental flavours.
The instanton corrections to the prepotential, the dual periods and the period matrix are calculated in a locus of special vacua possessing a $\mathbb{Z}_3$ symmetry. 
In a semi-classical expansion, we show that these observables are constrained by S-duality via a modular anomaly equation which takes the form of a recursion relation.
The solutions of the recursion relation are quasi-modular functions of $\Gamma_1 (3)$, which is a subgroup of the $S$-duality group and is also a congruence subgroup of $\text{SL}(2,\mathbb{Z})$.
}


\preprint{ }

\begin{document}
\section{Introduction}
\label{secn:intro}

Recently there has been a lot of progress in resumming instanton effects in conformal ${\mathcal N}=2$ supersymmetric gauge theories.%
\footnote{For a recent review on the exact results obtained in $\mathcal{N}=2$ theories we refer to \cite{Teschner:2014oja} and references therein.}
Most of this work has been done in the $\text{SU}(2)$ theory with four fundamental flavours
\cite{Huang:2011qx}\nocite{Billo:2013fi,Billo:2013jba}\,-\,\cite{Huang:2013eja}
and in the SU$(N)$ theories with an adjoint matter hypermultiplet, also called 
$\mathcal{N}=2^\star$ theories \cite{Billo:2014bja}\nocite{Billo':2015ria}\,-\,\cite{Billo':2015jta}.
In these analyses, the effective prepotential is organized as an 
expansion in inverse powers of the vacuum expectation values of the scalar field in the gauge
vector multiplet
and the non-perturbative contributions are computed using localization 
methods \`a la  Nekrasov \cite{Nekrasov:2002qd}\nocite{Nekrasov:2003rj,Bruzzo:2002xf}\,-\,\cite{Fucito:2004ry}.
The resulting expressions are functions of the ultraviolet (UV) gauge coupling $\tau_0$ or, equivalently, of the instanton counting parameter $q_0=\rme^{2\pi \ii \tau_0}$.
Quite remarkably, it has been shown that
these functions, after resumming the instanton corrections, 
become quasi-modular forms of the effective infrared (IR)
coupling of the massless theory on which the modular group 
$\Gamma=\mathrm{SL}(2,\mathbb{Z})$ faithfully acts in the standard fashion.

{For} $\mathcal{N}=2$ conformal SQCD theories with gauge groups of higher rank, progress along these lines
has been difficult because the S-duality group is no longer the full modular 
group $\Gamma$ but a discrete subgroup of $\mathrm{SL}(2,\mathbb{R})$.  
In this paper, we study the simplest of such theories, 
namely the $\mathcal{N}=2$ super Yang-Mills theory with gauge group $\text{SU}(3)$ and six 
fundamental flavours, which we describe in Section~\ref{secn:su3}. At a generic point on the Coulomb branch of its moduli space, one expects that the effective IR dynamics of the
theory is encoded by a Seiberg-Witten (SW) curve \cite{Seiberg:1994rs,Seiberg:1994aj} 
which, for the case at hand, is a surface of genus 2. We engineer such a curve in Section~\ref{secn:sw} using 
the $\text{NS}5-\text{D}4$ brane set-up in Type IIA string theory and its up-lift to 
M-theory \cite{Witten:1997sc}. The SW curve obtained in this way depends on the masses of the matter hypermultiplets, on the Coulomb branch parameters, and on the bare coupling constant $\tau_0$
through the parameter $q_0$. 

It is worth recalling that in the literature there have been several different proposals \cite{Hanany:1995na}\nocite{Argyres:1995wt, Minahan:1995er}\,-\,\cite{Minahan:1996ws}
for SW curves for superconformal theories, like the SU(3) model we are considering. These curves are always written in terms of an effective parameter called $\tau$.
Although these various proposals appear to be quite different, in \cite{Argyres:1999ty} it was shown that it is possible to map them onto each other by means of non-perturbative reparametrizations of the $\tau$ parameter, usually accompanied by a redefinition of the masses and of the vacuum expectation values in the Coulomb branch. However, these reparametrizations in general do not preserve 
global identifications on the $\tau$ space; in other words, the S-duality group one finds depends crucially on the choice of $\tau$ \cite{Argyres:1998bn}.

In this paper, in order to exhibit the duality properties in a transparent manner, we choose as $\tau$ the effective coupling constant of the massless theory in the so-called special vacuum \cite{Argyres:1999ty}. 
This is a locus on the Coulomb branch that possesses a $\mathbb{Z}_3$ symmetry and, in terms
of the classical vacuum expectation values $A_u$ of the adjoint scalar in the U(3) theory, it
corresponds to setting
\begin{equation}
A_u = \omega^{u-1} a \quad\mbox{for}~u=1,\,2,\,3\,,
\end{equation}
where $a$ is an arbitrary scale and $\omega$ is a cube root of unity. 
An added feature of the special vacuum is that, after decoupling a U(1) factor and reducing the gauge symmetry to SU(3), the period matrix $\Omega$ of the massless theory turns out to be proportional to the SU(3) Cartan matrix even when perturbative and instanton corrections are taken into account. Namely
\begin{equation}
\Omega = \tau\, \begin{pmatrix}
\,2 & -1\\
-1 & \,2
\end{pmatrix}~.
\label{cartan0}
\end{equation}
In the classical theory the prefactor is simply the bare gauge coupling $\tau_0$, but in the effective theory it
receives a perturbative correction at 1-loop and an infinite series of non-perturbative corrections due to instantons which can be explicitly computed using localization methods. Adding them up, one obtains the effective massless coupling $\tau$ on 
which the S-duality group acts 
faithfully, even when the masses for the hypermultiplets are turned on. 
It is interesting to note that the relation between $\tau_0$ and $\tau$ can be recovered 
by comparing the M-theory curve of Section~\ref{secn:sw} and the SW curve derived in \cite{Minahan:1995er, Minahan:1996ws}. Within the framework of \cite{Argyres:1999ty}, this turns out to be a particularly convenient choice of coordinates on the coupling constant space, in that it becomes possible to make contact with the results of the multi-instanton calculations in a straightforward and direct manner. 

Once the appropriate SW curve is singled out, we proceed to derive the action of S-duality on the period integrals of the curve. This action was already obtained in \cite{Argyres:1998bn}, but for the sake of completeness we briefly revisit this derivation in Section~\ref{secn:emduality}. A key feature of the duality transformations is that they leave the Cartan form (\ref{cartan0}) of the period matrix invariant while acting
on the $\tau$-parameter as
\begin{equation}
S~~~:~~~\tau ~\rightarrow~-\frac{1}{3\tau}\qquad \text{and}\qquad T~~~:~~~\tau~\rightarrow~ \tau+1~.
\end{equation}
The duality group generated by $S$ and $T$ admits as a subgroup
$\Gamma_1(3)$, which is also one of the congruence subgroups of $\Gamma$ \cite{Koblitz,Apostol}. 
One might therefore hope that there are observables that inherit quasi-modular properties under this subgroup and that it may be possible to resum the instanton contributions into quasi-modular forms, as was done for the SU$(2)$ case in \cite{Billo:2013jba,Billo:2013fi}. We explicitly realize this 
scenario in Section~\ref{secn:dual} where we show that S-duality indeed imposes strong constraints on the modular structure of several calculable quantities. 
In particular we find that the quantum part of the dual periods can be written in terms of quasi-modular forms of $\Gamma_1 (3)$ and that symmetry requirements 
from S-duality considerations imply a modular anomaly equation that takes the form of a recursion relation for the coefficients of the dual periods in the large-$a$ expansion. This recursion relation can then be used to calculate the higher order coefficients and verify that their instanton expansion completely agrees with direct computations 
performed using equivariant localization methods. We regard this agreement as a highly non-trivial consistency check of our results.

In Section~\ref{secn:concl} we present our conclusions and discuss some future perspectives, while 
several technical details are collected in the appendices.

\section{The $\mathcal{N}=2$ SU(3) theory with $N_f=6$}
\label{secn:su3}

We begin by reviewing the main features of the $\mathcal{N}=2$ superconformal theory with gauge group U(3) and six fundamental flavours. 

As usual it is convenient to combine the Yang-Mills coupling $g$ and the vacuum angle $\theta$ into the complex variable
\begin{equation}
\tau_0 = \frac{\theta}{2\pi}+\ii\,\frac{4\pi}{g^2}~,
\label{tau0}
\end{equation} 
and define the instanton counting parameter
\begin{equation}
q_0= \rme^{2\pi\ii\tau_0}~.
\label{q0}
\end{equation}
The prepotential of the theory is the sum of a classical term
$F_{\mathrm{class}}$, a perturbative 1-loop term $F_{\mathrm{1-loop}}$ and a non-perturbative piece $F_{\mathrm{inst}}$:
\begin{equation}
F=F_{\mathrm{class}}+F_{\mathrm{1-loop}}+F_{\mathrm{inst}}~.
\label{prep}
\end{equation}
The classical prepotential is
\begin{equation}
F_{\mathrm{class}}= \pi\ii\tau_0\sum_{u=1}^3 A_u^2
\label{fclass}
\end{equation}
with $A_u$ being the vacuum expectation value of the scalar field $A$ in the adjoint representation of U(3):
\begin{equation}
\langle A\rangle =\mathrm{diag}\big(A_1,A_2,A_3\big)~.
\label{vevU3}
\end{equation} 
The 1-loop term is independent of $\tau_0$ and is given by 
\begin{equation}
F_{\mathrm{1-loop}}= \sum_{u\not=v}\gamma(A_u-A_v)-
\sum_{u=1}^3\sum_{f=1}^6\gamma\big(A_u+m_f\big)\,,
\end{equation}
where 
\begin{equation}
\gamma(x)=-\frac{x^2}{4}\log\Big(\frac{x^2}{\Lambda^2}\Big)\,,
\end{equation}
with $\Lambda$ being an arbitrary mass scale.
Expanding $F_{\mathrm{1-loop}}$ for small masses, 
one obtains an expression in which the three vacuum expectation
values $A_u$ appear through the Casimirs of the U(3) gauge group
\begin{equation}
u_r=\sum_{u=1}^3 A_u^r\qquad\mbox{for}~r\in\{1,\cdots,3\}~,
\label{u}
\end{equation}
and the six fundamental masses $m_f$ through the Casimirs of the U(6) flavour symmetry group
\begin{equation}
T_\ell= \sum_{f=1}^6 m_f^\ell\qquad\mbox{for}~\ell\in\{1,\cdots,6\}~.
\label{casimirs}
\end{equation}
For simplicity in this paper we always consider mass configurations such that
\begin{equation}
T_1=m_1+m_2+m_3+m_4+m_5+m_6=0~.
\label{t1}
\end{equation}
Thus, the flavour symmetry group is reduced to SU(6).

The non-perturbative instanton part is computed using equivariant localization methods 
\cite{Nekrasov:2002qd}\nocite{Nekrasov:2003rj,Bruzzo:2002xf}\,-\,\cite{Fucito:2004ry}
(see also \cite{Billo:2012st} for technical details), 
and the result is of the form
\begin{equation}
F_{\mathrm{inst}}= \sum_{k=1}^\infty q_0^k\,f_k(u_r,T_\ell)\,,
\end{equation}
where the coefficients $f_k$ are rational functions of mass dimension 2 in which the Casimirs $T_\ell$ appear
only in the numerator. We have computed these functions up to $k=4$, but we do not
write them here since their explicit expressions are too lengthy%
\footnote{However, they can be made available upon request.}.

If the gauge group is SU(3) instead of U(3), we simply have to enforce the constraint
\begin{equation}
u_1=A_1+A_2+A_3=0
\label{su3}
\end{equation}
in all above formulas. We do this by setting
\begin{equation}
A_1=a_1~,\quad A_2=a_2-a_1~,\quad A_3=-a_2~,
\label{a12}
\end{equation}
which corresponds to choosing the two Cartan generators of SU(3) as $H_1=\mathrm{diag}(1,-1,0)$
and $H_2=\mathrm{diag}(0,1,-1)$. With this choice the classical prepotential (\ref{fclass}) becomes
\begin{equation}
F_{\mathrm{class}}= 2\pi\ii\tau_0\big(a_1^2-a_1 a_2+a_2^2\big)~,
\label{fclass1}
\end{equation}
from which we easily see that the classical period matrix
is
\begin{equation}
\big(\Omega_{\mathrm{class}}\big)_{ij} :=\frac{1}{2\pi\ii} \,\frac{\partial^2 F_{\mathrm{class}}}{\partial a_i \partial a_j}\, =\,\tau_0\,
\begin{pmatrix}
2 &-1\\
-1 & 2
\end{pmatrix}_{ij}~,
\label{omegacl}
\end{equation}
namely $\Omega_{\mathrm{class}}$ is proportional to the Cartan matrix of SU(3). 

When we add the perturbative and instanton corrections to the classical prepotential, this structure is generically lost. However there is a locus in the moduli space where the period matrix retains a simple form: it is 
the so-called \emph{special vacuum} \cite{Argyres:1999ty} which corresponds to setting
\begin{equation}
u_2=0~.
\label{sp}
\end{equation}
This condition can be achieved by taking a $\mathbb{Z}_3$-symmetric configuration 
of the $A_u$'s in the original U(3) theory, namely
\begin{equation}
A_u= \omega^{u-1}\,a\,,
\label{sp1}
\end{equation}
where $\omega= \rme^{\frac{2\pi\ii}{3}}$ and $a$ is an arbitrary scale. In terms of the SU(3) variables introduced in (\ref{a12}), the special vacuum corresponds to
\begin{equation}
a_1=a~,\quad a_2=-\omega^2\,a~.
\label{sp2}
\end{equation}
With this choice, the classical prepotential (\ref{fclass1}) vanishes, while the 1-loop term is%
\footnote{Here and in the following we always neglect all $a$-independent terms.}
\begin{equation}
F_{\mathrm{1-loop}}\Big|_{\mathrm{s.v.}} 
= 
\frac{T_2}{2}\log \Big(\frac{a^3}{\Lambda^3}\Big) +\frac{T_5}{20a^3}-\frac{T_8}{112a^6}+\frac{T_{11}}{330a^{9}} +\ldots
\label{Fpertsp}
\end{equation}
where the $T_\ell$'s with $\ell>6$ are defined as in (\ref{casimirs}) and can be written as polynomials 
of the independent lower Casimirs.
Adding the instanton corrections, the full prepotential in the special vacuum becomes
\begin{equation}
F\,\Big|_{\mathrm{s.v.}}=
\frac{T_2}{2}\log \Big(\frac{a^3}{\Lambda^3}\Big)+\sum_{n=1}^\infty
\frac{\widehat{h}_n(q_0,T_\ell)}{n\,a^{3n}}
\label{Fq0}
\end{equation}
where the first few coefficients $\widehat{h}_n$
are reported, up to four instantons, in Appendix~\ref{secn:appa} (see in particular (\ref{hathnq0})).
The period matrix $\Omega$ in the special vacuum takes a rather simple form; 
defining the matrices
\begin{equation}
\label{CBBdagger}
\mathcal{C} = \begin{pmatrix}
2 & -1 \\
-1 & 2
\end{pmatrix} ~,\quad
\mathcal{B} =\begin{pmatrix}
\omega^2 & \omega \\
\omega & 1
\end{pmatrix}
 \quad\mbox{and}\quad
\mathcal{B}^{\dagger} =
\begin{pmatrix}
\omega & \omega^2 \\
\omega^2 & 1
\end{pmatrix}~,
\end{equation}
one finds:
\begin{eqnarray}
\Omega  &=&~\frac{1}{2\pi \ii}\left( 2\pi \ii\tau_0 -\ii\pi-\log27+\frac{4q_0}{9}+\frac{14q_0^2}{81}+\frac{1948 q_0^3}{19683}+\cdots\right){\mathcal C}\label{Omegasp}\\
&& - \frac{1}{2\pi \ii}\sum_{n=0}^{\infty} \left[ \left( \frac{3n+1}{a^{3n+2}} \widehat{h}_n(q_0,T_\ell) \right) \mathcal{B} + \left( \frac{3n+2}{a^{3n+3}} \widehat{g}_n(q_0,T_\ell) \right) \mathcal{C} + \left( \frac{3n+3}{a^{3n+4}} \widehat{k}_n(q_0,T_\ell) \right) \mathcal{B}^\dagger \right].
\nonumber
\end{eqnarray}
The $\widehat{h}_n(q_0,T_\ell)$'s are the same functions that appear in the prepotential (\ref{Fq0}), 
while $\widehat{g}_n(q_0,T_\ell)$ and $\widehat{k}_n(q_0,T_\ell)$ can be reconstructed from the
expressions given in Appendix~\ref{secn:appa} (see (\ref{gnq}) and (\ref{knq})). Looking at them, 
it is easy to realize that the coefficients of $\mathcal{B}$ and ${\mathcal B}^{\dagger}$ 
vanish if $T_2=T_4=T_5=0$. 

Prompted by this observation, we conclude that turning on only
$T_{3}$ and $T_6$ yields a period matrix proportional to $\cC$ with the constant of proportionality
playing the role of an effective IR coupling constant, 
even after taking into account the $1$-loop and instanton contributions. 
In terms of the bare masses, this is equivalent to making a $\mathbb{Z}_3$-symmetric choice, for instance
\begin{equation}
(m_1, m_2, m_3) = (m, m\,\omega, m\,\omega^2) \quad \text{and} \quad 
(m_4, m_5, m_6) = (m', m'\omega, m'\omega^2)~,
\label{masses}
\end{equation}
for which the non-zero Casimirs $T_3$ and $T_6$ become
\begin{equation}
T_3 = 3\,(m^3+{m'}^3)\qquad\text{and}\qquad T_6 = 3\,(m^6+{m'}^6)~.
\label{t3t6}
\end{equation}
With this mass configuration, the period matrix maintains the same simple form (\ref{omegacl}) of the classical theory, namely 
\begin{equation}
\Omega = \widetilde{\tau} \, \mathcal{C}~,
\label{omegaC}
\end{equation}
where the effective IR coupling $\widetilde{\tau}$ is
\begin{eqnarray}
2\pi \ii \widetilde{\tau} &=& 2\pi \ii \tau_0 -\ii\pi-\log27+\frac{4q_0}{9}+\frac{14q_0^2}{81}+\frac{1948 q_0^3}{19683}+\cdots\phantom{\Big|} \nonumber\\
&&
+\,\frac{T_3}{a^3}\Big(\frac{1}{3}-\frac{2q_0}{27}-\frac{8q_0^2}{243}-\frac{398q_0^3}{19683}+\cdots \Big)
-\frac{T_6}{a^6}\Big(\frac{1}{6}+\frac{5q_0}{27}+\frac{5q_0^2}{243}+\frac{5q_0^3}{19683} +\cdots\Big)\phantom{\Bigg|}
\nonumber\\
&&
+\,\frac{T_3^2}{a^6}\Big(\frac{5q_0}{81}+\frac{10q_0^2}{729}+\frac{95q_0^3}{19683}+\cdots\Big)+\cdots ~.
\label{tautilde}
\end{eqnarray}
A quantity of interest, which will play a crucial role in later sections, is the effective IR coupling of the \emph{massless} theory. Notice that even when all masses are zero there is a non-perturbative renormalization of the bare UV coupling $\tau_0$ to an effective IR coupling%
\footnote{This is different from the $\mathcal{N}=2^\star$ theories 
\cite{Billo:2013fi,Billo:2013jba,Billo:2014bja} where for zero mass one recovers the maximal 
$\mathcal{N}=4$ supersymmetry which implies that the coupling constant does not run, neither
perturbatively nor non-perturbatively.}. 
We will denote this coupling by $\tau$ which, as we can see from (\ref{tautilde}) with
$T_3=T_6=0$,  is given by
\begin{equation}
\label{tau}
2\pi \ii \tau =  2\pi \ii \tau_0 -\ii \pi-\log27+\frac{4q_0}{9}+\frac{14q_0^2}{81}+\frac{1948 q_0^3}{19683}+\cdots ~.
\end{equation}
This relation can be inverted and, exponentiating both sides we find
\begin{equation}
q_0 = -27\,q \left(1+12 q+90q^2+ 508q^3+\cdots \right) ~,
\label{UVIRinst}
\end{equation}
where $q=\rme^{2\pi \ii \tau}$. 
As we shall see in the following sections, the massless IR coupling  $\tau$ plays an important role because 
it is the natural parameter that transforms faithfully under the S-duality transformations of the effective theory \cite{Minahan:1997fi}.

\section{The Seiberg-Witten curve}
\label{secn:sw}

In this section we derive the SW curve for the SU(3) gauge theory with $N_f=6$ flavours discussed before. 
We engineer such a theory using the $\text{NS}5-\text{D}4$ brane set up of Type IIA string theory
\cite{Witten:1997sc}, and closely follow the notations and conventions of \cite{Ashok:2015gfa}.

\subsection{Derivation from M-theory}
Our construction begins with a collection of $\text{NS}5$ and $\text{D}4$ branes, arranged as indicated in Tab.~\ref{Table1}.
\begin{table}[ht]
\begin{center}
\begin{tabular}{|c|c|c|c|c|c|c|c|c|c|c||c|}
\hline
\phantom{\Big|}  & $x^0$ & $x^1$ & $x^2$ & $x^3$ & $x^4$ & $x^5$ & $x^6$ & $x^7$ & $x^8$ & $x^9$ & $x^{10}$\\
\hline
$\text{NS}5$ branes & $-$ & $ -$ &$ -$ & $-$ &$-$ & $-$ & $\cdot$ & $\cdot$ & $\cdot$ & $\cdot$ &$\cdot$
\\
\hline
$\text{D}4$ branes  & $-$ &$ -$ &$ -$ & $-$ &$\cdot $ & $\cdot$ & $-$ & $\cdot$ & $\cdot$& $\cdot$  &$-$
\\
\hline
\end{tabular}
\caption{Type IIA brane configuration: the symbols $-$ and $\cdot$ denote longitudinal and transverse directions respectively; the last column refers to the eleventh dimension after the M-theory uplift.\label{Table1}}
\end{center}
\end{table}
The first four directions $\{x^0,x^1,x^2,x^3\}$ are longitudinal for both kinds of branes and 
span the spacetime $\mathbb{R}^{1,3}$ where the gauge theory is defined. 
After compactifying the $x^5$ direction on a circle $S^1$ of radius $R_5$, we uplift the system
to M-theory by introducing a compact eleventh coordinate $x^{10}$ with radius $R_{10}$. 
We then minimize the world-volume of the resulting $\text{M}5$ branes and in 
doing so, obtain the SW curve for a $5$-dimensional $\mathcal{N}=1$ gauge theory 
defined in $\mathbb{R}^{1,3} \times S^1$ which takes the form of 
a $2$-dimensional surface inside the space parameterized by the coordinates $\{x^4,x^5, x^6,x^{10} \}$.
To get the curve for the 4-dimensional $\mathcal{N}=2$ theory, we first T-dualize along $x^5$ and 
then take the limit of small (dual) radius. Thus, in terms of the dual circumference
\begin{equation}
\beta= \frac{2\pi\alpha'}{R_5} ~,
\label{beta}
\end{equation}
the $4$-dimensional limit corresponds to $\beta \rightarrow 0$. 

Let us now give some details. To engineer the conformal $\text{SU}(3)$ gauge theory, we consider 
the following configuration (see Fig.~1):
\begin{itemize}
	\item Two $\text{NS}5$ branes separated by finite distances along the $x^6$ direction, labeled $\text{NS}5_1$ and $\text{NS}5_2$,
	\item Three semi-infinite $\text{D}4$ branes on $\text{NS}5_1$ and three semi-infinite $\text{D4}$ branes on the $\text{NS}5_2$,
	\item Three finite $\text{D}4$ branes suspended between $\text{NS}5_1$ and $\text{NS}5_2$.
\end{itemize}
\begin{figure}[ht!]
\centering
\includegraphics[width=140mm]{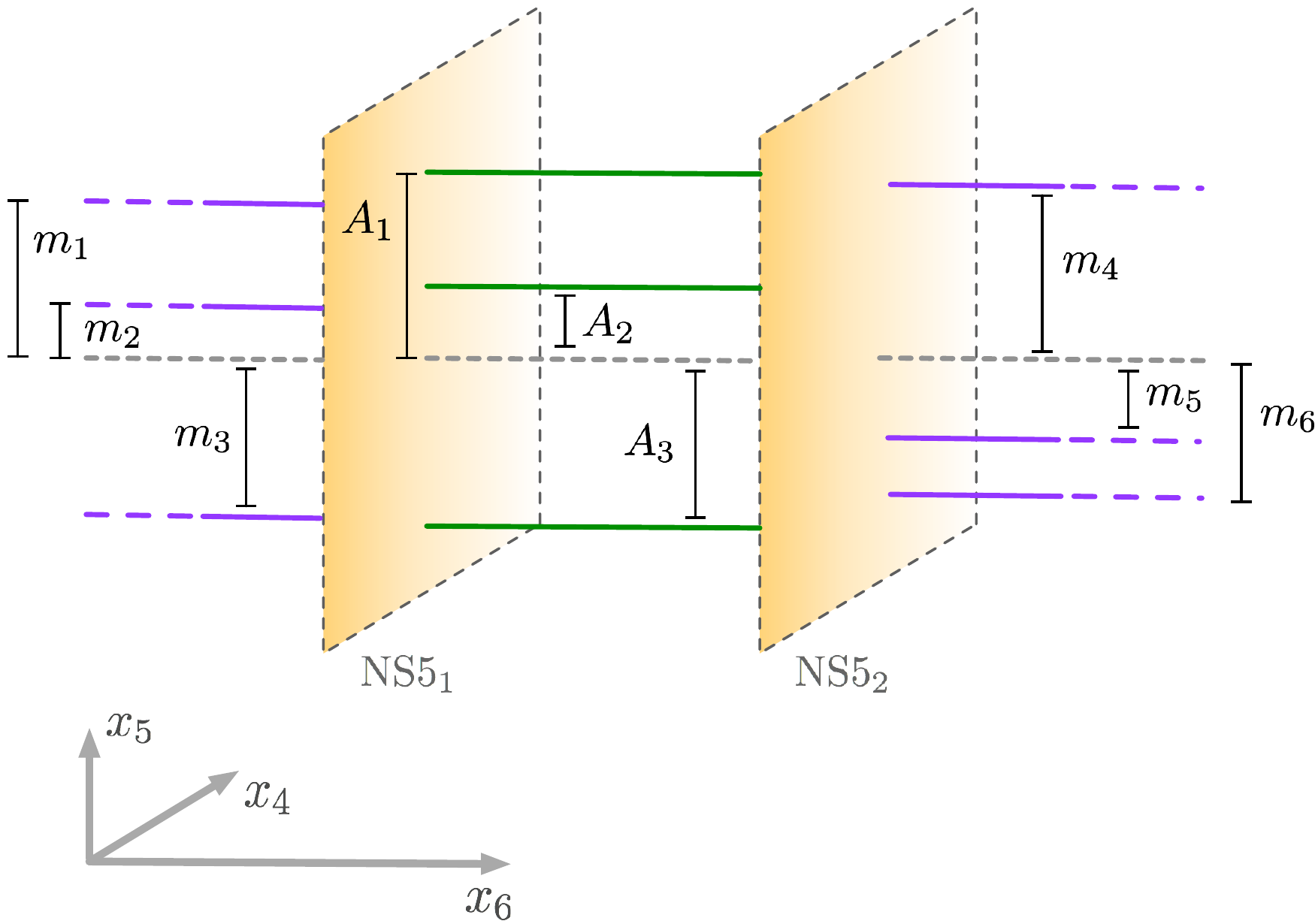}
\caption{NS5 and D4 brane set up for the conformal SU$(3)$ theory. \label{overflow}}
\end{figure}
The brane configuration is best described in terms of the complex combinations
\begin{equation}
x^4+\ii \, x^5 \equiv 2\pi\alpha' v \qquad \text{and} \qquad x^6+\ii \, x^{10} \equiv s \, ,
\label{vs}
\end{equation}
or their exponentials
\begin{equation}
w \equiv \rme^{\frac{2\pi\alpha' v}{R_5}} = \rme^{\beta v} \qquad \text{and} \qquad
t \equiv \rme^{\frac{s}{R_{10}}} \, ,
\label{wt}
\end{equation}
which are single-valued under integer shifts of $x^5$ and $x^{10}$ along the respective circumferences.

The SW curve for the T-dual 5-dimensional gauge theory takes the form
of a polynomial equation of degree 2 in $t$, since there are two NS5 branes, and of degree 3 in $w$,
since there are three suspended D4 branes. Such a polynomial equation can be
written in two distinct ways; the first is \cite{Ashok:2015gfa}
\begin{equation}
{\mathcal C}_{1}: \quad 
w^3\,\big(t-t^{(\infty)}_{1}\big)\big(t-t^{(\infty)}_{2}\big) +w^2\,Q _2(t)+w\,Q_1(t)
+d' \big(t-t^{(0)}_{1}\big)\big(t-t^{(0)}_{2}\big) = 0 ~.
\end{equation}
Here, $Q_1(t)$ and $Q_2(t)$ are quadratic polynomials in $t$, $d'$ is an arbitrary coefficient,
while $t^{(0)}_{i}$ and $t^{(\infty)}_i$  are the asymptotic positions of the two $\text{NS}5$ branes as $w\to 0$ and $w\to \infty$ respectively. These are given by
\begin{equation}
\begin{aligned}
t^{(\infty)}_{1} &= C^{(1)}\sqrt{\frac{\widetilde{m}_{1}\,\widetilde{m}_{2}\,\widetilde{m}_{3}} {\widetilde{A}_{1}\,\widetilde{A}_{2}\,\widetilde{A}_{3}} }~, \qquad
t^{(0)}_{1} = C^{(1)}\sqrt{\frac{\widetilde{A}_{1}\,\widetilde{A}_{2}\,\widetilde{A}_{3}}{\widetilde{m}_{1}\,\widetilde{m}_{2}\,\widetilde{m}_{3}} }~,\\
t^{(\infty)}_{2} &= C^{(2)}\sqrt{\frac{\widetilde{A}_{1}\,\widetilde{A}_{2}\,\widetilde{A}_{3}} {\widetilde{m}_{4}\,\widetilde{m}_{5}\,\widetilde{m}_{6}} }~, \qquad
t^{(0)}_{2} = C^{(2)}\sqrt{\frac{\widetilde{m}_{4}\,\widetilde{m}_{5}\,\widetilde{m}_{6}}{\widetilde{A}_{1}\,\widetilde{A}_{2}\,\widetilde{A}_{3}}}
\end{aligned}
\label{curvec1}
\end{equation}
where $C^{(1)}$ and $C^{(2)}$ are constants and the tilded variables are defined as
\begin{equation}
\widetilde{X} = \rme^{\beta X}
\end{equation}
for any $X$. In particular the tilded variables of (\ref{curvec1}) correspond to the positions of the D4 branes in the $(x^4,x^5)$-plane which represent the three vacuum expectation values $A_u$ 
of the adjoint scalar field and the six masses $m_f$ of the U(3) gauge theory, as described in 
the previous section. Since we are interested in the SU(3) theory, we impose the constraint
(\ref{su3}) which corresponds to setting the center of mass of the colour D4 branes to the origin.

The second form of the SW curve is \cite{Ashok:2015gfa}
\begin{equation}
\label{dualcurve}
{\mathcal C}_{2}: \quad (w-\widetilde{m}_{1})(w-\widetilde{m}_{2}) (w-\widetilde{m}_{3})\,t^2 + P(w)\, t + d(w-\widetilde{m}_{4})(w-\widetilde{m}_{5}) (w-\widetilde{m}_{6})=0 ~,
\end{equation}
where $P(w)$ is a cubic polynomial and $d$ is an arbitrary coefficient. This structure follows from the fact that when $t\to 0$ and $t\to\infty$ there are three D4 flavour branes at $w=(\widetilde{m}_1,\widetilde{m}_2,\widetilde{m}_3)$ and $w=(\widetilde{m}_4,\widetilde{m}_5,\widetilde{m}_6)$, respectively.

Equating (\ref{curvec1}) and (\ref{dualcurve}), we can fix all but two coefficients in terms of the physical parameters of the gauge theory and get
\begin{multline}
(w-\widetilde{m}_{1})(w-\widetilde{m}_{2}) (w-\widetilde{m}_{3})\,t^2 \cr
- \Big[\big(t^{(\infty)}_1+t^{(\infty)}_2\big)\,w^3+c_2\,w^2+c_1\,w-\widetilde{m}_1
\widetilde{m}_2 \widetilde{m}_3\big(t^{(0)}_1+t^{(0)}_2\big)\Big]\,t \cr
+t^{(\infty)}_1\,t^{(\infty)}_2 (w-\widetilde{m}_{4})(w-\widetilde{m}_{5}) (w-\widetilde{m}_{6}) = 0 ~~~
\end{multline}
where the two undetermined coefficients, $c_1$ and $c_2$, parametrize the Coulomb branch of the SU(3) gauge theory. According to \cite{Witten:1997sc}, the difference in asymptotic positions of the NS5 branes is
related to the bare gauge coupling constant; to be precise, one has
\begin{equation}
q_0= \frac{C^{(1)}}{C^{(2)}} ~.
\end{equation}

Now we can take the four dimensional limit $\beta\to 0$. Skipping several intermediate steps which are similar to those described in \cite{Ashok:2015gfa}, we find that the 
four dimensional SW curve for the SU(3) superconformal theory is
\begin{equation}
\label{SWoldM}
(v-m_1)(v-m_2)(v-m_3) \,t^2 - \left ((1+q_0)v^3 -U_2\, v -U_3 \right)t +q_0(v-m_4)(v-m_5)(v-m_6) =0~.
\end{equation}
Here we have used the fact that $T_1=\sum_f m_f=0$ and have redefined the coefficients $c_1$ and $c_2$ 
to absorb all terms linear in $t$ and proportional to $v$ or independent of $v$ into the new parameters $U_2$ and $U_3$. These have been normalized in such a way that
\begin{equation}
U_2\,\big|_{q_0\to 0}=u_2\qquad\text{and}\qquad U_3\,\big|_{q_0\to 0}=u_3~.
\end{equation}
In the special vacuum and in the $\mathbb{Z}_3$-symmetric mass configuration (\ref{masses}), the SW curve
(\ref{SWoldM}) further simplifies to
\begin{equation}
(v^3-m^3)\,t^2- \left ((1+q_0)v^3 -U_3 \right)t +q_0(v^3-{m'}^3)=0~.
\label{SWsimpl}
\end{equation}

\subsection{Relating the UV and IR parameters}
The SW curve (\ref{SWoldM}) is written in terms of the UV parameters of the SU(3) gauge theory; in particular
it explicitly depends on $q_0$ which is the instanton counting parameter 
used in the Nekrasov analysis of the non-perturbative prepotential. However, it would be useful 
to have a description in terms of the effective IR parameters and couplings since these are the ones which faithfully transform under the duality group of the conformal gauge theory. 

To do this let us first rewrite the curve (\ref{SWoldM}) in a different form \cite{Gaiotto:2009hg}. Define the  monic polynomial 
\begin{equation}
\mathcal{P}(v) = v^3 - \frac{U_2}{1+q_0}\,v -\frac{U_3}{1+q_0}
\label{calP}
\end{equation}
and rescale $t$ into $(1+q_0)\,t$, so that (\ref{SWoldM}) becomes
\begin{equation}
\label{SWoldM1}
(v-m_1)(v-m_2)(v-m_3) \,t^2 - \mathcal{P}(v)\,t +\frac{q_0}{(1+q_0)^2}(v-m_4)(v-m_5)(v-m_6) =0~.
\end{equation}
If we shift away the term linear in $t$ and introduce
\begin{equation}
y = 2 (v-m_1)(v-m_2)(v-m_3) \left(t-\frac{\mathcal{P}(v)}{2(v-m_1)(v-m_2)(v-m_3)} \right) ~,
\end{equation}
we obtain
\begin{equation}
\label{SWMnew}
y^2 = \mathcal{P}^2(v) -\frac{4q_0}{(1+q_0)^2}\,\mathcal{Q}(v)
\end{equation}
where
\begin{equation}
\mathcal{Q}(v)=\prod_{f=1}^6(v-m_f)~.
\label{calQ}
\end{equation}
In this way we manage to cast the M-theory curve for our superconformal $\text{SU}(3)$ gauge theory 
into a hyperelliptic form which makes comparison with the earlier literature \cite{Hanany:1995na, Argyres:1995wt, Minahan:1995er, Minahan:1996ws} straightforward. 
In particular, consider the curve given in Eq.~(4.12) of \cite{Minahan:1996ws}. 
This curve is written in terms of the IR coupling $\tau$ of the massless theory in the special vacuum and, after a shift and rescaling, it takes the form 
\begin{equation}
\label{MNcurve}
y^2 = \mathcal{P}^2(v) - h(\tau)\,\mathcal{Q}(v)
\end{equation}
where 
\begin{equation}
h(\tau)=\left(\frac{f_-^2(\tau)-f_+^2(\tau)}{f_-^2(\tau)}\right)~.
\label{htau}
\end{equation}
Here $f_{\pm}(\tau)$ are modular forms of $\Gamma_1(3)$ with weight $3$, given by
\begin{equation}
\label{fpm}
f_{\pm}(\tau) = \left(\frac{\eta^3(\tau)}{\eta(3\tau)}\right)^3\pm \left(\frac{3\eta^3(3\tau)}{\eta(\tau)}\right)^3
\end{equation}
with $\eta(\tau)$ being the Dedekind $\eta$-function \cite{Minahan:1996ws} (see also
\cite{Koblitz,Apostol}). We refer to Appendix~\ref{secn:modular} for
the properties of the modular functions and a brief review of $\Gamma_1(3)$ which, as we 
show later, arises as a subgroup of the S-duality group of the SU(3) superconformal theory. 

The two curves (\ref{SWMnew}) and (\ref{MNcurve}) become identical if
\begin{equation}
\frac{f_+(\tau)}{f_-(\tau)} = \frac{1-q_0}{1+q_0}~,
\end{equation}
which implies
\begin{equation}
\label{UVIRcurve}
q_0 = -27\left(\frac{\eta(3\tau)}{\eta(\tau)} \right)^{12}~.
\end{equation}
This relation has already been obtained in \cite{Billo:2012st} using different arguments related to the holographic description of the effective gauge coupling. 
Fourier expanding the $\eta$-functions in powers of $q=\rme^{2\pi\ii\tau}$, we obtain 
\begin{equation}
\label{qvsQ}
q_0 = -27\,q \left(1+12 q+90q^2+ 508q^3+\cdots \right)
\end{equation}
which exactly agrees with (\ref{UVIRinst}). The fact that the parameter $\tau$ appearing in the 
modular functions of the curve (\ref{MNcurve}) is related to the effective IR coupling of the massless
theory in the special vacuum (and hence to the period matrix $\Omega$) provides 
a nice geometrical interpretation, while the fact that 
the UV/IR relation (\ref{UVIRcurve}) is in perfect agreement with the explicit multi-instanton calculations 
gives a strong confirmation on the validity of the whole picture.

\section{Electric-magnetic duality}
\label{secn:emduality}
The SW curve (\ref{MNcurve}) represents a genus 2 surface on which we can choose 
a canonical homology basis of cycles $\{\alpha_i\}$ and $\{\beta_j\}$ (with $i,j=1,2$) 
such that $\alpha_i\cap\beta_j=\delta_{ij}$. On these cycles the $\text{Sp}(4,\mathbb{Z})$ electric-magnetic duality group acts in the standard fashion.
We denote an element of this group by
\begin{equation}
M=\begin{pmatrix}
A & B\\
C & D
\end{pmatrix} \in \text{Sp}(4,\mathbb{Z})~,
\end{equation}
where $A$, $B$, $C$ and $D$ are $2\times 2$ matrices satisfying the relations 
\begin{align}
A^TC - C^T A = 0~, \qquad B^TD - D^TB = 0 ~, \qquad\text{and}\qquad A^T D - C^TB = \one_2 ~.
\end{align}
Under Sp$(4,\mathbb{Z})$ the homology cycles transform as
\begin{equation}
\begin{pmatrix}
\beta\\
\alpha
\end{pmatrix} ~\rightarrow ~ M \cdot \begin{pmatrix} 
\beta\\
\alpha
\end{pmatrix}=
\begin{pmatrix}
A \,\beta+B\,\alpha\\
C\,\beta+D\,\alpha
\end{pmatrix}~,
\label{atrasnf}
\end{equation}
while the period matrix transforms as
\begin{equation}
\label{EMOmega}
\Omega \longrightarrow \big(A\,\Omega+B\big)\big(C\,\Omega +D\big)^{-1} ~.
\end{equation}

Given the SW differential $\lambda$, the period integrals over the $\alpha_i$- and 
$\beta_j$-cycles are denoted as
\begin{equation}
a_i= \oint_{\alpha_i} \lambda \qquad \text{and} \qquad a_j^\text{D} = \oint_{\beta_j} \lambda~.
\end{equation}
The $\alpha$-periods represent the vacuum expectation values of the adjoint scalar field; that is why we used the same notation as in Section~\ref{secn:su3}. The $\beta$-periods, instead, 
represent their dual variables which are related to the prepotential $F$ according to
\begin{equation}
a_i^{\text{D}} =\frac{1}{2\pi \ii}\,\frac{\partial F}{\partial a_i}~.
\label{adual} 
\end{equation}
{From} the transformation rules (\ref{atrasnf}) of the cycles under $\text{Sp}(4,\mathbb{Z})$, it immediately
follows that the period integrals transform according to
\begin{equation}
\label{Sponaduals}
\begin{pmatrix}
a^{\text{D}}\\
a
\end{pmatrix} ~\rightarrow~ 
M\cdot
\begin{pmatrix}
a^{\text{D}}\\
a 
\end{pmatrix} =  \begin{pmatrix}
A \,a^{\text{D}}+B\,a\\
C\,a^{\text{D}}+D\,a
\end{pmatrix}~.
\end{equation}

\subsection{The S-duality group}

The S-duality groups of ${\mathcal N}=2$ conformal SQCD theories have been studied in 
\cite{Minahan:1997fi,Argyres:1998bn}. Here we will briefly review the analysis of \cite{Argyres:1998bn} adapting it to our $\text{SU}(3)$ theory. 

Referring to the SW curve (\ref{MNcurve}), the total moduli space has the structure of a fiber bundle in which the $h$-plane is the base and the Coulomb branch is the fiber. 
In the conventional SW analysis of non-scale invariant theories \cite{Seiberg:1994rs}, the monodromies around 
the singularities of the Coulomb branch generate the low energy electric-magnetic duality group. In
scale invariant theories, like the one we are considering, there are in addition isolated ``special'' singularities in the space of coupling constants where the entire Coulomb branch degenerates. 
Monodromies around these special points in the $h$-plane are \textit{defined} 
to be the generators of the S-duality group \cite{Argyres:1998bn}. 

\begin{figure}[t]
\centering
\includegraphics[width=60mm]{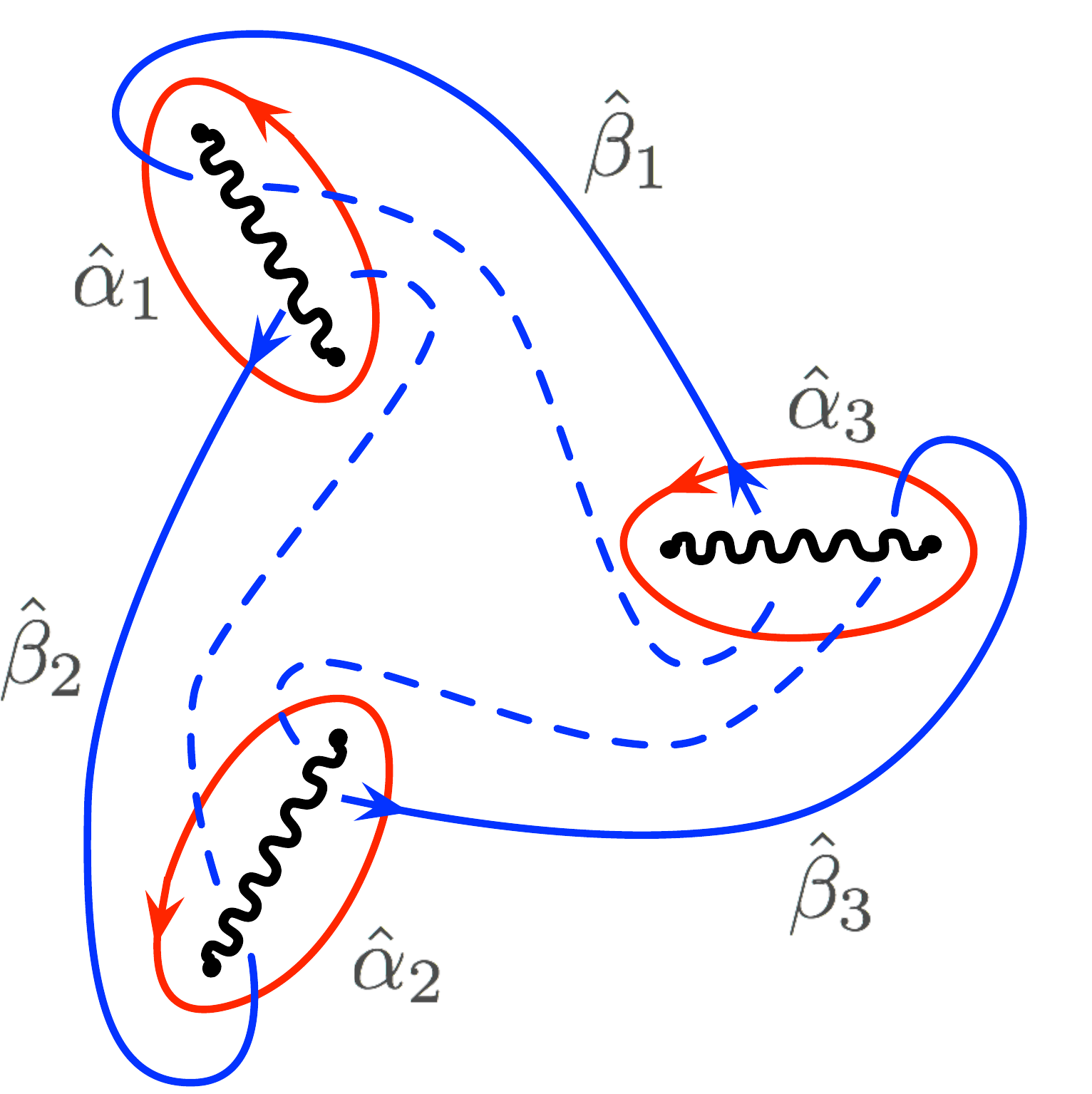}
\caption{The $\hat{\alpha}$ and $\hat{\beta}$ cycles for the U(3) SW curve. \label{fig:cycles}}
\end{figure}

The discriminant locus of the SW curve (\ref{MNcurve}) is proportional to 
\begin{equation}
\Delta = h^3(h-1)^3 ~.
\end{equation}
Thus, the curve is singular at $h=0,1$ and $\infty$. In the special vacuum, these are the only singularities
around which the curve degenerates \cite{Argyres:1998bn}.
To discuss how S-duality is realized on our SW curve, let us temporarily return to a U(3) set-up and choose
a $\mathbb{Z}_3$-symmetric set of cycles $\{\hat{\alpha}_u\}$ and $\{\hat{\beta}_v\}$ 
($u,v=1,2,3$) as shown
in Fig.~\ref{fig:cycles}, whose intersection matrix is
\begin{equation}
\Big(\hat{\alpha}_u\cap\hat{\beta}_v\Big) = \begin{pmatrix}
1 & -1 &  0\cr
0 & 1  & -1\cr
-1& 0  & 1
\end{pmatrix}~.
\label{intersec}
\end{equation}
These cycles are not linearly independent since
\begin{equation}
\sum_{u=1}^3 \hat{\alpha}_u =0 \qquad\mbox{and}\qquad \sum_{v=1}^3\hat{\beta}_v = 0~.
\end{equation}

Under the monodromy $h\rightarrow e^{2\pi \ii}h$, one can check that the branch points are rotated such that the cycles are transformed as follows%
\footnote{Here we use the convention that the indices $u$ and $v$ are defined modulo 3 and that the
hatted operators act on the U$(3)$ basis of cycles while the unhatted ones correspond 
to Sp$(4,\mathbb{Z})$ matrices which are obtained by restricting the action 
of the hatted operators to the SU$(3)$ basis which we introduce later.}
\begin{equation}
\hat T\quad : \quad \begin{cases}
\hat{\alpha}_u ~\rightarrow~ \hat{\alpha}_u ~ , \\
\hat{\beta}_v ~\rightarrow~ \hat{\beta}_v+\hat{\alpha}_v-\hat{\alpha}_{v-1} ~.
\end{cases}
\label{Tu3}
\end{equation}
Under $(h-1)\rightarrow e^{2 \pi \ii}(h-1)$ , we obtain the monodromy $\hat S \hat T^{-1}$, which acts on the cycles as
\begin{equation}
\hat S \,\hat T^{-1} \quad : \quad \begin{cases}
\hat{\alpha}_u ~\rightarrow~ \hat{\beta}_u-\hat{\alpha}_u+\hat{\alpha}_{u-1} ~, \\
\hat{\beta}_v ~\rightarrow ~\hat{\alpha}_{v-1} ~.
\end{cases}
\end{equation}
It then follows that the $\hat S$ transformation, which corresponds to the monodromy around $h\to\infty$,  is given by
\begin{equation}
\hat S \quad : \quad \begin{cases}
\hat{\alpha}_u ~\rightarrow~ \hat{\beta}_u ~, \\
\hat{\beta}_v ~\rightarrow~ \hat{\alpha}_{v-1} ~.
\end{cases}
\label{Su3}
\end{equation}

We have already remarked that the Riemann surface is invariant under a $\mathbb{Z}_3$
rotation of the cycles generated by
\begin{equation}
\label{defPhi}
\hat\Phi \quad : \quad \begin{cases}
\hat{\alpha}_u ~\rightarrow~\hat{\alpha}_{u-1} ~, \\
\hat{\beta}_v ~\rightarrow ~\hat{\beta}_{v-1} ~.
\end{cases}
\end{equation}
Therefore, instead of the transformation $\hat S$ we can equivalently consider the transformation 
obtained by composing it with $\Phi$, {\it{i.e.}} we define a modified $\hat S$ transformation 
\begin{equation}
\label{defShat}
\hat S  ~\rightarrow  ~\hat\Phi\cdot \hat S\quad : \quad \begin{cases}
\hat{\alpha}_u ~\rightarrow~ \hat{\beta}_{u-1} ~, \\
\hat{\beta}_v ~\rightarrow~ \hat{\alpha}_{v+1} ~.
\end{cases}
\end{equation}    
For simplicity $\hat\Phi\cdot \hat S$ will be renamed $\hat S$.
The new duality transformations satisfy
\begin{equation}
\label{SThatrel}
\hat S^2=1 \qquad\text{and}\qquad (\hat S\hat T)^{6} = 1~.
\end{equation} 

Let us now return to the SU(3) theory and choose the independent $\alpha$- and $\beta$-cycles as 
\begin{align}
\alpha_1 &= \hat{\alpha}_1\qquad \text{and}\qquad \alpha_2 =-\hat{\alpha}_3~, \cr
\beta_1 &= -\hat{\beta}_2 \,\quad\text{and}\qquad \beta_2 = -\hat{\beta}_3 ~.
\label{cyclesab}
\end{align}
Using (\ref{intersec}), it is easy to verify that $\alpha_i \cap \beta_j=\delta_{ij}$, as it should be for a canonical homology basis. The restrictions of the $\hat S$ and $\hat T$ transformations to the 
SU$(3)$ basis (\ref{cyclesab}) follow directly from the transformation 
rules (\ref{Tu3}) and (\ref{defShat}) in the parent 
U(3) theory. Representing them as $4\times 4$ matrices acting on the 4-vector 
{\small{$\begin{pmatrix}\beta\\ \alpha \end{pmatrix}$}}, 
we have
\begin{equation}
T=\begin{pmatrix}
1&0&2&-1\cr
0&1&-1&2\cr
0&0&1&0\cr
0&0&0&1
\end{pmatrix}\qquad \text{and}\qquad
S= 
\begin{pmatrix}
0&0&0&1\cr
0&0&-1&0\cr
0&-1&0&0\cr
1&0&0&0
\end{pmatrix}~.
\label{STmat}
\end{equation}

Acting projectively as described in (\ref{EMOmega}) on the period matrix $\Omega=\tau \, \mathcal{C}$  these matrices leave its matrix structure invariant, while the prefactor $\tau$ changes as follows
\begin{equation}
\label{dualtau}
T\quad:~\tau~ \longrightarrow~ \tau+1 \qquad\text{and}\qquad S\quad:~\tau~ \longrightarrow -\frac{1}{3\tau}~. 
\end{equation}
These transformations satisfy  the relations
\begin{equation}
S^2=1 \qquad\text{and}\qquad (ST)^{6} = 1~,
\end{equation} 
just as  their U(3) parents $\hat T,\hat S$ in (\ref{SThatrel}).
They generate a discrete subgroup of $\text{SL}(2,\mathbb{R})$ which is the duality group of the theory.

As we observed earlier, this duality group has a natural action also on the periods $a_i$ and their dual ones 
$a_j^{\mathrm{D}}$ which is inherited from the action of $T$ and $S$ on the $\alpha_i$ and $\beta_j$ cycles.
In particular under the $S$-transformation in \eqref{STmat}, we have
\begin{equation}
\label{Sdualperiods}
S\quad:~\begin{pmatrix}
a_{1}^{\text{D}}\\a_{2}^{\text{D}}\\a_1\\a_2 
\end{pmatrix}~\longrightarrow ~\begin{pmatrix}
a_{2}\\-a_1\\-a_{2}^{\text{D}}\\a_{1}^{\text{D}} 
\end{pmatrix}~.
\end{equation}
Note that under this transformation the $a_i$'s and the $a_j^{\text{D}}$ are exchanged (up to signs), 
as expected for an $S$-transformation. Note also that 
the transformations (\ref{dualtau}) leave invariant the function $h(\tau)$ in (\ref{htau}) that appears 
in the SW curve (\ref{MNcurve}) of the SU(3) theory, as can be checked using the modular
properties of the Dedekind $\eta$-functions reported in Appendix~\ref{secn:modular}.

\subsubsection{An important subgroup: $\Gamma_1(3)$}

An important subgroup of the full duality group, which will play an central role in what follows, is the one generated by $T$ (as in (\ref{STmat})) and $S'$ where \cite{Minahan:1997fi}
\begin{equation}
S' = S\,T\,S^{-1} 
= 
\begin{pmatrix}
1&0&0& 0 \\
0&1&0&0\\
-2&-1&1&0\\
-1&-2&0&1
\end{pmatrix}~.
\label{S'}
\end{equation}
This $S'$ rotates the periods into each other according to
\begin{equation}
\label{Sprimedualperiods}
S'\quad:~ 
\begin{pmatrix}
a_{1}^{\text{D}}\\a_{2}^{\text{D}}\\a_1\\a_2 
\end{pmatrix}~\longrightarrow~ 
\begin{pmatrix}
a_{1}^{\text{D}}\\
a_{2}^{\text{D}}\\
a_{1}-2a_{1}^{\text{D}}-a_{2}^{\text{D}}\\
a_2 -a_{1}^{\text{D}}-2a_{2}^{\text{D}}
\end{pmatrix}~,
\end{equation}
while it acts on the coupling $\tau$ as
\begin{equation}
S'\quad:~\tau~ \longrightarrow~ \frac{\tau}{1-3\tau}~.
\end{equation}
$T$ and $S'$ generate the group $\Gamma_1(3)$, which is a congruence subgroup of the modular 
group SL$(2,\mathbb{Z})$:
\begin{equation}
\Gamma_1(3) = \Bigg\{
\begin{pmatrix}
~a~&~b~ \\
~c~&~d~
\end{pmatrix} \in \mathrm{SL}(2,\mathbb{Z})~: ~c=0~\,\mbox{mod}~3~,~a=d=1~\,\mbox{mod}~3
\Bigg\}~.
\label{gamma13}
\end{equation}
As we have already mentioned, the function $h(\tau)$ appearing in the SW curve
(\ref{MNcurve}) is written in terms of modular forms of $\Gamma_1(3)$. In what follows, we will exploit this symmetry
and in particular, using the above action of S-duality, we will show that the dual periods of 
the $\text{SU}(3)$ theory in the special vacuum can be expanded in terms of quasi-modular 
functions of $\Gamma_1(3)$.

\section{The dual periods and S-duality}
\label{secn:dual}

In order to study the S-duality properties of the superconformal $\text{SU}(3)$ theory, 
we have to calculate the dual periods ${a}_i^{\text{D}}$ in the special vacuum. 
This is done by deriving the prepotential $F$ and then setting the scalars to their vacuum expectation values of the special vacuum:
\begin{equation}
a_i^{\text{D}} = \frac{1}{ 2\pi \ii}\left. \frac{\partial F}{\partial a_i}\right|_{\text{s.v.}}~.
\end{equation}
In the special vacuum, there is only one independent pair of dual variables, which we choose 
to be $(a_1, a_{2}^{\text{D}})$. This follows from the S-duality rules derived in (\ref{Sdualperiods}).
Thus, from now on we use the notation $a_1\equiv a$ and $a_{2}^{\text{D}}\equiv a^{\text{D}}$.

We consider the classical, 1-loop and instanton contributions separately. {From} the prepotential 
(\ref{fclass1}) we find that the classical contribution is%
\footnote{For completeness, we point out that in the special vacuum the period $a_{1}^{\text{D}}$ is classically proportional to $a_{2}^{\text{D}}$. Even when the instanton corrections are taken into account, the following relation holds in the special vacuum:
\begin{equation*}
a_{1}^{\text{D}}\big|_{\text{s.v.}}= \frac{1}{\omega}\, a_{2}^{\text{D}}\big|_{\text{s.v.}} ~.
\end{equation*}
}
\begin{equation}
a^{\text{D}}_{\text{class}} =\ii \sqrt{3} \tau_0\, a~.
\label{adualcl}
\end{equation}
The $1$-loop contribution to the dual period in the special vacuum is given by
\begin{equation}
a^{\text{D}}_{1\text{-loop}} = \frac{\sqrt{3}}{2\pi}\left[ (-\ii\pi - \log27)a-\frac{T_3}{6a^2} +\frac{T_6}{30a^5}+\frac{T_3(T_3^2-9T_6)}{1296a^8}
+\cdots\right]~,
\label{adualpert}
\end{equation}
while the instanton corrections, which are a power series in $q_0$, can be calculated using the 
techniques of equivariant localization. 

These results, however, are not very illuminating. Things change significantly if we write the dual 
period using the massless effective coupling $\tau$, which transforms faithfully under the S-duality group. To do so, we have use the UV/IR relation (\ref{qvsQ}). Once this is done, we find that $a^{\text{D}}$
takes the form
\begin{equation}
\label{a2Dexp}
a^{\text{D}} = \ii \sqrt{3}\tau\, a
+\frac{ \sqrt{3}}{2\pi} \sum_{n=0}^{\infty} \frac{g_n(\tau,T_\ell)}{a^{3n+2}} ~,
\end{equation}
where the first few $g_n$'s have the following $q$-expansions
\begin{subequations}
\begin{align}
g_0 &= -\frac{T_3}{6} \left(1+6q+ 6q^3
+ \cdots \right) ~, \\
g_1 &= \frac{T_3^2}{3}\left(q+6q^2+3q^3+\cdots\right) 
+\frac{T_6}{30} \left(1-30q-270q^2-570q^3+ \cdots \right) ~, \\
g_2&=\frac{T_3^3}{1296} \left(1-756q^2-336q^3+ \cdots \right)
-\frac{T_3T_6}{144}\left(1-252q^2+672q^3+ \cdots \right)~,\\
g_3 &= -\frac{T_3^4}{7128}\left(1+1386 q^2-11088 q^3+\cdots\right) + 
	\frac{T_3^2 T_6}{1188}\left(1+1386 q^2-5808 q^3+\cdots\right) \notag\\
	&~~~ + \frac{T_6^2}{792} \left(1-1386 q^2+528 q^3+\cdots\right) ~, 
	\label{g3q}\\
g_4 &= \frac{T_3^5}{68040}\left(1+87360 q^3+\cdots\right)
	-\frac{T_3^3 T_6}{27}\left(208 q^3 +\cdots \right)\notag\\
	&~~~ - \frac{T_3 T_6^2}{1512} \left(1-17472 q^3+\cdots\right)
	\label{g4q}
\end{align}
\label{gnexp}
\end{subequations}
when only $T_3$ and $T_6$ are turned on. In what follows we will show that these coefficients $g_n$'s, 
which appear also in front of $\cC$ in the period matrix, are in fact quasi-modular 
functions of $\Gamma_1(3)$ with weight $3n+1$. 

\subsection{S-duality and modularity}

In Section~\ref{secn:emduality} we have already derived how S-duality acts on the periods and their duals. In order to see the consequences of this when we are in the special vacuum, let us consider the combination 
\begin{equation}
\label{Xexp}
X = a^{\text{D}}-\ii\sqrt{3}\tau a =\frac{\sqrt{3}}{2\pi}\sum_n \frac{g_n(\tau, T_\ell)}{a^{3n+2}}
\end{equation}
and study its transformation properties under both $S$ and $S'$ using the Sp$(4,\mathbb{Z})$ matrices defined in \eqref{Sdualperiods} and \eqref{Sprimedualperiods} respectively. 

\subsubsection{The $S'$ transformation}
Under the $S'$-transformation, we find
\begin{equation}
\begin{aligned}
a^\text{D} & ~\rightarrow~  a^\text{D} ~, \\
a&~\rightarrow~ (a+\ii\sqrt{3}\, a^\text{D})~.
\end{aligned}
\end{equation}
{From} this it follows that $X$ transforms as 
\begin{equation}
S'(X) =\frac{X}{(1-3\tau)}~.
\label{SX1}
\end{equation}
If we act with $S'$ on the right hand side of (\ref{Xexp}), we find
\begin{equation}
S'(X) =\frac{\sqrt{3}}{2\pi}\sum_n \frac{S'\big(g_n(\tau,T_\ell)\big)}{a^{3n+2}(1-3\tau)^{3n+2}}
\left(1-\frac{3}{2\pi \ii(1-3\tau)}\sum_m \frac{g_m}{a^{3m+3}} \right)^{-3n-2}~.
\label{SX2}
\end{equation}
Equating (\ref{SX1}) and (\ref{SX2}), we deduce that, to leading order, the $g_n$'s must behave as 
modular forms of $\Gamma_1(3)$ with weight $3n+1$. Indeed,
\begin{equation}
S'\big(g_n(\tau, T_\ell)\big) := g_n\left(\frac{\tau}{1-3\tau}, S'(T_\ell)\right) = (1-3\tau)^{3n+1} 
\big((g_n(\tau, T_\ell) + \cdots \big) ~.
\label{S1X}
\end{equation}
The modular forms of $\Gamma_1(3)$ form a ring generated by two elements: 
$f_1(\tau)=\big(f_+(\tau)\big)^{1/3}$, which has weight 1, and $f_-(\tau)$, which has weight $3$. Here
$f_\pm(\tau)$ are the functions defined in (\ref{fpm}). As we will see in more detail later, the fact that the above equation holds only at leading order and that there are corrections as the ellipses in (\ref{S1X}) indicate, implies that the $g_n$'s are actually \emph{quasi}-modular functions of $\Gamma_1(3)$. By this we mean that the $g_n$'s are polynomials in $f_1(\tau)$, $f_-(\tau)$ and $E_2(\tau)$, where $E_2(\tau)$ is the second Eisenstein series of weight 2 defined by
\begin{equation}
E_2(\tau) = 1-24\sum_{n=1}^{\infty}\sigma_1(n)q^n = 1-24q-72q^2-96q^3+\cdots
\label{E2}
\end{equation}
with $\sigma_1(n)$ being the sum of divisors of $n$. $E_2$ is not a modular form owing to its anomalous behaviour under the $\text{SL}(2,\mathbb{Z})$ transformations:
\begin{equation}
E_2\left(\frac{a\tau+b}{c\tau+d}\right) = (c\tau+d)^2 E_2(\tau) + \frac{6c}{\ii\pi}(c\tau+d)  ~.
\end{equation}
The modular forms of $\Gamma_1 (3)$ can be divided into two classes according to their parity under 
the $S$-transformation $\tau\,\to\,-\ft{1}{3\tau}$. Indeed one has (see Appendix~\ref{secn:modular})
\begin{equation}
\begin{aligned}
f_1\Big(\!-\frac{1}{3\tau}\Big) &=  -\, (\ii \sqrt{3}\tau)\, f_1(\tau) ~,\\
f_- \Big(\!-\frac{1}{3\tau}\Big) &= +\, (\ii\sqrt{3}\tau)^{3} \, f_-(\tau) ~,
\end{aligned}
\end{equation}
so that we can assign $S$-parity $(+1)$ to $f_-$, and $S$-parity $(-1)$ to $f_+$. Since all modular forms of
$\Gamma_1(3)$ are generated by these basis elements, the above assignments are enough to specify the
$S$-parity of any modular form of $\Gamma_1(3)$. Furthermore, one can show that the combination
\begin{equation}
\widetilde{E}_2(\tau) = E_2(\tau) + f_1^2(\tau)
\label{tildeE2}
\end{equation}
has $S$-parity $(-1)$ since, as we show in Appendix~\ref{secn:modular},
\begin{equation}
\label{EtildeStransform}
\widetilde{E}_2 \Big(\!-\frac{1}{3\tau} \Big) = - (\ii\sqrt{3}\tau)^2 \Big(\widetilde{E}_2(\tau)  
+ \frac{6}{\ii \pi\tau}\Big)~.
\end{equation}
We are now adequately prepared to understand how $X$ transforms under S-duality.

\subsubsection{The $S$ transformation}
The $S$-transformation (\ref{Sdualperiods}) implies that the periods in the special vacuum transform as
\begin{equation}
\begin{aligned}
a^{\text{D}} &~\rightarrow~ -a ~,\\
a &~\rightarrow~ -a^{\text{D}} ~.
\end{aligned}
\end{equation}
Therefore, from (\ref{Xexp}) we have
\begin{equation}
S(X) = \frac{X}{\ii\sqrt{3} \tau} ~.
\label{SX11}
\end{equation}
Acting with $S$ on the right hand side of (\ref{Xexp}) we get
\begin{equation}
S(X)=  \frac{\sqrt{3}}{2\pi}\sum_{n=0}^{\infty}(-1)^n 
\frac{S\big(g_n(\tau, T_\ell)\big)}{(\ii\sqrt{3}\tau a)^{3n+2}}\left( 1+\frac{1}{2\pi \ii \tau}\sum_m \frac{g_m}{a^{3m+3}}\right)^{-3n-2}~.
\label{SX111}
\end{equation}
Equating (\ref{SX11}) and (\ref{SX111}), at leading order we find that
\begin{equation}
g_n \Big(\!-\frac{1}{3\tau}, S(T_\ell)\Big) = (-1)^n\left(\ii\sqrt{3}\tau\right)^{3n+1}
\big( g_n(\tau, T_\ell) + \cdots \big) ~.
\end{equation}
{From} the behaviour of $g_n$ we can now infer that, to leading order, the $g_n$'s are modular forms of $\Gamma_1(3)$ with parity $(-1)^n$. The sub-leading terms will modify this to the statement that the $g_n$'s are \emph{quasi-}modular forms of $\Gamma_1(3)$, with definite parity.

Let us now give some details. In the simplest case of $g_0$, there are no subleading corrections to the $S$ and $S'$ transformations. Our previous argument tells us that $g_0$ is a modular form of $\Gamma_1(3)$ with
weight 1 and $S$-parity $+1$. As mentioned above, there is a unique modular form of weight 1, namely
\begin{equation}
f_1(\tau) =\big(f_+(\tau)\big)^{\frac{1}{3}} = 1+ 6q + 6q^2 + 6q^3+ \cdots ~.
\end{equation}
While this perfectly agrees with the $q$-expansion in \eqref{gnexp}, $f_1$ has $S$-parity $-1$.  
Therefore, in order for the coefficients to match, the $T_\ell$'s must also transform under the $S$-transformation according to
\begin{equation}
S(T_3) = -T_3 \qquad\text{and}\qquad S(T_6) = T_6 ~.
\label{ST3T6}
\end{equation}
This completes the identification of $g_0$ as a modular form of weight 1 and $S$-parity $+1$, namely
\begin{equation}
g_0(\tau, T_\ell) = -\frac{T_3}{6}\, f_1(\tau)~.
\label{g0fin}
\end{equation}
One could repeat this analysis for the other $g_n$'s. However, it is more convenient and efficient
to exploit a recursion relation satisfied by these coefficients, to which we now turn.

\subsection{Recursion relation}

Before stating the general result, let us work out the explicit dual transformation properties 
of the lower $g_n$'s. Since the behaviour under $S$ and $S'$ are very similar, we will only 
exhibit the transformation rules under $S$ which are a bit simpler to discuss. 
We begin by comparing the coefficient of the term proportional to $1/a^5$ in the two expressions (\ref{SX11}) and (\ref{SX111}) for $S(X)$. {From} the first one we have
\begin{equation}
S(X)\Big|_{{1}/{a^5}}= \frac{1}{2\pi\ii\tau}\,g_1~,
\end{equation}
while from the second we get
\begin{equation}
S(X)\Big|_{{1}/{a^5}}= -\frac{\sqrt{3}}{2\pi}\,\Bigg[\frac{S(g_1)}{(\ii\sqrt{3}\tau)^5}+\frac{S(g_0)}{(\ii\sqrt{3}\tau)^2}\,\frac{g_0}{\pi \ii \tau}\Bigg]~.
\end{equation}
Using the fact that $S(g_0)=(\ii\sqrt{3}\tau)\,g_0$, as one can see from (\ref{g0fin}), the above two expressions imply
\begin{equation}
S(g_1)=-(\ii\sqrt{3}\tau)^4\,\Big(g_1+\frac{g_0^2}{\pi\ii\tau}\Big)~.
\label{Sg1}
\end{equation}
The presence of the $g_0^2$ term clearly shows that $g_1$ cannot be simply a modular form; in fact to compensate for its presence, $g_1$ must be a quasi-modular form, as we have already anticipated.
The only quasi-modular form is the second Eisenstein series $E_2$ or its modified version $\widetilde{E}_2$
defined in (\ref{tildeE2}). To stress the fact that $\widetilde{E}_2$ should appear in the expression for $g_1$,
we temporarily modify the notation and, in place of $g_1(\tau,T_\ell)$, we write
\begin{equation}
g_1(\widetilde{E}_2,T_3,T_6)~.
\end{equation}
We instead continue to leave implicit the dependence on the other modular forms to avoid clutter in the formulas.
Then, using the known $S$-transformation of $\widetilde{E}_2$ (see \eqref{EtildeStransform}), and demanding 
that $g_1$ has modular weight $4$ and $S$-parity $(-1)$, we have
\begin{equation}
\label{Sg12}
\begin{aligned}
S(g_1) &:= g_1\big(S(\widetilde{E}_2),S(T_3),S(T_6)\big)=g_1\left(\widetilde{E}_2(-\ft{1}{3\tau}),-T_3,T_6\right)\\
&\,=-(\ii\sqrt{3}\tau)^4\, g_1\Big(\widetilde{E}_2+\ft{6}{\pi\ii\tau},-T_3,T_6 \Big) \\
&\,=-(\ii\sqrt{3}\tau)^4\,\left[ g_1\big(\widetilde{E}_2,-T_3,T_6 \big) +\frac{6}{\pi\ii\tau}
\,\frac{\partial}{\partial\widetilde{E}_2}g_1\big(\widetilde{E}_2,-T_3,T_6 \big) +\cdots\right]\\
\end{aligned}
\end{equation} 
where we have used (\ref{EtildeStransform}) and (\ref{ST3T6}) and the dots stand for terms of higher order in $1/\tau$ which are proportional to higher $\widetilde{E}_2$-derivatives. Comparing with (\ref{Sg1}),
we deduce that $g_1$ must be an even function of $T_3$ and that, in order to match the $1/\tau$ terms,
the following relation must hold
\begin{equation}
\label{recursiong1}
\frac{\partial g_1}{\partial \widetilde{E}_2} = \frac{1}{6}\,g_0^2~.
\end{equation}
Integrating with respect to $\widetilde{E}_2$, we find that
\begin{equation}
g_1 = \frac{T_3^2}{216}f_1^2\widetilde{E}_2 + g_1^{(0)}
\end{equation}
where $g_1^{(0)}$ is a modular form of $\Gamma_1(3)$ with weight $4$ and $S$-parity $-1$.
As shown in Tab.~2 of Appendix~\ref{secn:modular}, there is only one such modular form, namely $f_1\,f_-$,
and thus $g_1^{(0)}$ must be proportional to it. The proportionality coefficient can be fixed by comparing with
the perturbative value of the $1/a^5$ term of $X$, which is $T_6/30$ as one can deduce from (\ref{adualpert}). The final result for $g_1$ is then
\begin{equation}
g_1 =\frac{T_3^2}{216}\big(f_1^2\widetilde{E}_2-2f_1\, f_-\big) +\frac{T_6}{30}\,f_1\,f_-~.
\end{equation}
Notice that this is an even function of $T_3$ as it should be, and that it is linear in $\widetilde{E}_2$. Thus
in (\ref{Sg12}) there are no higher order terms in $1/\tau$, in perfect agreement with (\ref{Sg1}).

By expanding the modular forms in powers of $q$ we can obtain the multi-instanton contributions to $g_1$ and
check that they perfectly agree with the explicit results reported in (\ref{gnexp}) and obtained
with the equivariant localization methods. The fact that by using the differential equation (\ref{recursiong1}) and \emph{only} the 1-loop result it is possible to get the entire tail of instanton corrections and the fact
that these are in complete agreement with the explicit multi-instanton calculations are a very strong and
highly non-trivial consistency check of our analysis.

The process we have described can be repeated order by order in $n$ for the higher coefficients $g_n$, and the final result can be compactly stated in the form of a recursion relation:
\begin{equation}
\frac{\partial g_n}{\partial \widetilde{E}_2} =\frac{(3n+1)}{24}\sum_{\ell<n}\, g_\ell\, g_{n-\ell-1} ~.
\label{recursiongn}
\end{equation}
Using this relation, one can determine the quasi-modular parts of $g_n$. The remaining terms are fixed in terms of modular 
functions of $\Gamma_1(3)$ with definite $S$-parity by comparing the first few coefficients of the $q$-expansion. For example, at the next order we find
\begin{equation}
g_2= -\frac{T_3^3}{1296}\Big(\frac{7f_1^3 \widetilde{E}_2^2}{24}-\frac{7f_1^2f_-\widetilde{E}_2}{6} +\frac{11f_1^7}{18} -\frac{7f_1 f_-^2}{9}\Big)-\frac{T_3 T_6}{144} \Big(\frac{7 f_1^2 f_-\widetilde{E}_2 }{15}+\frac{3f_1^7}{10}-\frac{7f_1 f_-^2}{30}\Big)~.
\label{g2fin}
\end{equation}
To fix the coefficients of the modular forms $f_1^7$ and $f_1\,f_-^2$ we have matched the 1-loop and the
1-instanton terms, while the higher instanton corrections are predicted by the $q$-expansion of the
right hand side. Again the perfect agreement with the explicit results from localization is a highly non-trivial
consistency check.

Using this method we also computed $g_3$ and $g_4$. Their expressions are given in 
Appendix~\ref{secn:appgn}. 

\section{Summary and discussion}
\label{secn:concl}

In this work we have studied the ${\mathcal N}=2$ supersymmetric conformal SU$(3)$ gauge theory with six fundamental flavours restricted to a ${\mathbb Z}_3$-symmetric locus of its Coulomb branch. The main results are based on the calculation of the prepotential, the dual periods and the period matrix in this special vacuum. These include both one-loop and instanton contributions (up to instanton number 4) to these observables, for general mass configurations. 

For a restricted $\mathbb{Z}_3$-invariant mass configuration, the period matrix was shown to be proportional to the Cartan matrix of SU$(3)$:
\begin{equation}
\Omega = \tau(q_0) 
\begin{pmatrix}
2 & -1\\
-1 &2
\end{pmatrix} ~.
\end{equation}
The proportionality constant, in the massless limit, defines an effective $\tau$-parameter,
and we showed that the S-duality group acts faithfully on it. 
We further obtained the S-duality action on the relevant 
period integrals $(a, a^{\text{D}})$ in the special vacuum. Combining this with the general form 
of the dual periods in the (large-$a$) semi-classical expansion,
\begin{equation}
a^{\text{D}} = \ii\sqrt{3}\tau\, a+\frac{\sqrt{3}}{2\pi}\sum_n\frac{g_n(\tau, T_k)}{a^{3n+2}}~,
\end{equation}
we showed that the transformation of $\tau$ and the period integrals $(a, a_D)$ under the S-duality group constrains the coefficients $g_n(\tau)$ to be quasi-modular forms of $\Gamma_1(3)$. Effectively, this amounts to resumming all instanton contributions at a given order in the large-$a$ expansion. The quasi-modular part of $g_n$, namely the part that depends on the second Eisenstein series $E_2(\tau)$ or
its extension $\widetilde{E}_2(\tau)$ defined in (\ref{tildeE2}), is completely determined by means of the following recursion relation: 
\begin{equation}
\frac{\partial g_n}{\partial \widetilde{E}_2} =\frac{(3n+1)}{24}\sum_{\ell<n}\, g_\ell\, g_{n-\ell-1} ~.
\end{equation}
This determines $g_n$ up to modular functions of $\Gamma_1(3)$ with a given weight $3n+1$. 

Of particular note in our derivation of quasi-modularity is the role played by the full S-duality group and not just the subgroup $\Gamma_1(3)$. 
There are numerous modular forms of $\Gamma_1(3)$ for a given weight (see Tab.~2 in Appendix B) and naively, one would have expected that the knowledge of the instanton expansion to a very high order were necessary to fix the modular part completely. 
However, the constraint from the $S$-transformation only allows for those modular forms of $\Gamma_1(3)$ that have 
fixed $S$-parity to appear in the expansion of the dual periods. Taking this into account, it proves sufficient to use only the perturbative and $1$-instanton contribution to completely
determine the first few $g_n$ coefficients. Remarkably, the higher order instanton calculations are completely consistent with the resulting expansion in terms of quasi-modular functions. 

It would be interesting to extend our analysis away from the special vacuum. 
At a generic point in the moduli space, the period matrix is no longer proportional to the Cartan matrix.
Given that the S-duality group acts as a subgroup of Sp$(4,\mathbb{Z})$, we would expect that duality would constrain calculable quantities to have an expansion in terms of genus-2 theta functions. Similarly, turning on the $\Omega$-deformation parameters would also take us away from the special vacuum and it would be interesting to see whether there is any sort of quasi-modular behaviour of the prepotential or dual periods in such cases. 

Once our analysis is extended to a generic point in the Coulomb moduli space, it would be interesting to explore whether our results could be useful in studying the infinite coupling point in the moduli space of couplings \cite{Argyres:2007cn}. Since the exact dependence on the coupling constant is known, one could take the coupling to be close to the infinite coupling limit and directly derive properties of the dual theories. This could give us useful information about non-Lagrangian theories, such as the superconformal theories with $E_n$ symmetry \cite{Minahan:1996fg,Minahan:1996cj}, that would be difficult to obtain otherwise. 

Finally, the S-duality group of superconformal quivers has been studied in great detail in \cite{Gaiotto:2009we}. Furthermore the AGT-W correspondence relates the partition function of the $\Omega$-deformed gauge theory to chiral conformal blocks in two dimensional Toda CFT \cite{Alday:2009aq, Wyllard:2009hg}. In the SU(2) case, recursion relations that encapsulate the quasi-modular structure of the prepotential have been obtained in \cite{KashaniPoor:2012wb, Kashani-Poor:2013oza, Kashani-Poor:2014mua} by studying null-vector decoupling equations in Liouville theory. It would be interesting to see if one can recover the recursion relations derived in this paper by studying similar equations in Toda theory. 

\vskip 1.5cm
\noindent {\large {\bf Acknowledgments}}
\vskip 0.2cm
We thank Carlo Angelantonj, Soumyadeep Bhattacharya, Francesco Fucito, Laurent Gallot, Hossein Parsa Ghorbani, Dileep Jatkar and Renjan John for several discussions.

The work of M.B., M.F. and A.L. is partially supported  by the Compagnia di San Paolo 
contract ``MAST: Modern Applications of String Theory'' TO-Call3-2012-0088.

\vskip 1cm

\appendix
\section{Observables in the special vacuum}
\label{secn:appa}

In this appendix we collect the explicit expressions for some of the observables of the SU(3) theory in the
special vacuum obtained using equivariant localization. All observables are first calculated at a generic point in the U$(3)$ moduli space after which the special vacuum constraint \eqref{sp1} is imposed. 

\subsection*{The prepotential coefficients}
In the special vacuum the prepotential takes the form
\begin{equation}
F\,\Big|_{\mathrm{s.v.}}=
-\widehat{h}_0\, \log \Big(\frac{a^3}{\Lambda^3}\Big)+\sum_{n=1}^\infty
\frac{\widehat{h}_n(q_0,T_\ell)}{n\,a^{3n}}~.
\label{Fq0app}
\end{equation}
Up to four instantons, the first few coefficients $\widehat{h}_n(q_0,T_\ell)$ are
\begin{subequations}
\begin{align}
\widehat{h}_0 &= -\frac{T_2}{2} ~, \\
\widehat{h}_1 &= \frac{T_5}{131220} \left(6561+8748 q_0+3402 q_0^2+2028 q_0^3+1412 q_0^4+\cdots\right)\notag \\
	&~~~-\frac{T_2 T_3}{39366} \left(2187q_0+972 q_0^2+603 q_0^3+428 q_0^4+\cdots\right) ~, \\
\widehat{h}_2 &= -\frac{T_2^4}{5878656} \left(2187-2268 q_0^2-2464 q_0^3-2156 q_0^4+\cdots\right)\notag \\
	&~~~+ \frac{T_2^2 T_4}{1469664} \left(6561-6804 q_0^2-7392 q_0^3-6692 q_0^4+\cdots\right)\notag \\
	&~~~- \frac{T_4^2}{1469664} \left(6561+6804 q_0^2+4704 q_0^3+3500 q_0^4+\cdots\right) \notag
	\\
	&~~~- \frac{T_3 T_5}{229635} \left(2187-567 q_0^2-742 q_0^3-700 q_0^4+\cdots\right)\notag \\
	&~~~+\frac{T_2T_3^2}{551124} \left(2187-3402 q_0^2-3472 q_0^3-3080 q_0^4+\cdots\right) \notag\\
	&~~~- \frac{T_2T_6}{183708} \left(2187-2268 q_0^2-2128 q_0^3-1820 q_0^4+\cdots\right)~.
\end{align}
\label{hathnq0}
\end{subequations}
As they stand these expressions are not very illuminating. Indeed, it is more useful to replace the instanton counting parameter $q_0$ and rewrite them in terms of the effective coupling $q$ using the UV/IR relation 
(\ref{UVIRinst}). Denoting the new coefficients obtained in this way as $h(q,T_\ell)$, we have
\begin{subequations}
\begin{align}
h_0 &= -\frac{T_2}{2}~,\\
h_1 &= \frac{T_5}{20}\left(1-36q-54q^2-252q^3 -468 q^4+\cdots\right) \notag\\
&~~~+ 
\frac{3T_2T_3}{2}\left(q+3q^3 +4q^4+\cdots \right)~,\\
h_2 &= -\frac{T_2^4}{2688}\left(1-756q^2+4032q^3+29484q^4+\cdots\right)  \notag\\
&~~~
+\frac{T_2^2T_4}{224}\left(1-756q^2+4032q^3+11340q^4+\cdots\right)\notag\\
&~~~-\frac{T_4^2}{224}\left(1+756q^2+4032q^3+20412q^4+\cdots\right) \notag\\
&~~~
-\frac{T_3T_5}{105}\left(1-189q^2+2142q^3+9072q^4+\cdots\right) \notag\\
&~~~+ \frac{T_2T_3^2}{252}\left(1-1134q^2+4032q^3+9072q^4+\cdots\right) \notag\\
&~~~
-\frac{T_6T_2}{84}\left(1-756q^2+1008q^3+2268q^4+\cdots\right)~.
\end{align}
\label{hnq}
\end{subequations}
We observe that for the mass configuration (\ref{masses}) for which only the Casimirs $T_3$ and $T_6$ are
non-zero, all coefficients $\widehat{h}_n$ (or $h_n$) vanish.

\subsection*{The dual period coefficients}
We present the results for the dual period $a^{\text{D}}:= \frac{1}{2\pi\ii}\frac{\partial F}{\partial a_2}\Big|_{\mathrm{s.v.}}$ in the special vacuum. This is initially obtained as an expansion in the microscopic coupling $q_0$. As for the prepotential, the result takes a much simpler form when reexpressed in terms of the effective coupling $q$:
\begin{equation}
a^{\text{D}} = \ii\sqrt{3}\tau\, a+\frac{\sqrt{3}}{2\pi} 
\sum_{n=0}^{\infty} \Big[\frac{h_n(q,T_\ell)}{a^{3n+1}} + \frac{g_n(q,T_\ell)}
{a^{3n+2}}\Big]
\end{equation}
where the functions $h_n(q,T_\ell)$ are the same as in (\ref{hnq}) while the first few
functions $g_n(q,T_\ell)$ up to 4 instantons are given by
\begin{subequations}
\begin{align}
g_0 &= -\frac{T_3}{6} \left(1+6q+ 6q^3 + 6q^4+\cdots\right) ~,\\
g_1 &= -\frac{T_2^3}{8}\left(q+27q^3-224q^4+\cdots\right) \notag\\
&~~~+\left(\frac{T_3^2}{3}+\frac{3T_2T_4}{4}\right)\left(q+6q^2+3q^3+28q^4+\cdots\right)\notag\\
&~~~+\frac{T_6}{30}\left(1-30q-270q^2-570q^3-2190q^4+\cdots\right)~,\\
g_2&=\frac{T_2^3T_3}{576}\left(1-468q^2-5952q^3+4914q^4+\cdots\right)\notag\\
&~~~+\frac{T_3^3}{1296}\left(1-756q^2-336q^3+35154q^4+\cdots\right)\notag\\
&~~~+\frac{3T_2T_3T_4}{4}\left(q^2+26q^3+186q^4+\cdots\right)\notag\\
&~~~-\frac{T_2^2T_5}{320}\left(1-396q^2-6528q^3+1026q^4+\cdots\right)\notag\\
&~~~-\frac{T_4T_5}{160}\left(1+396q^2+5856q^3+35586q^4+\cdots\right)\notag\\
&~~~-\frac{T_3T_6}{144}\left(1-252q^2+672q^3+20034q^4+\cdots\right) ~.
\end{align}
\label{gnq}
\end{subequations}
When we set $T_2=T_4=T_5=0$, we obtain the result quoted in \eqref{gnexp}.

\subsection*{The period matrix coefficients}
In the special vacuum the period matrix can be put in the following form
\begin{equation}
\begin{aligned}
\Omega &=\tau\, {\mathcal C} - \frac{1}{2\pi \ii}\sum_{n=0}^{\infty} \left[ \left( \frac{3n+1}{a^{3n+2}} h_n(q,T_\ell) \right)\! \mathcal{B} +\! \left( \frac{3n+2}{a^{3n+3}} g_n(q,T_\ell) \right)\! \mathcal{C} +\! \left( \frac{3n+3}{a^{3n+4}} k_n(q,T_\ell) \right) \!\mathcal{B}^\dagger \right]
\end{aligned}
\label{Omegaspapp}
\end{equation}
where $\mathcal{C}$ is the Cartan matrix of SU(3), and the matrices $\mathcal{B}$ and $\mathcal{B}^{\,\dagger}$ are defined in (\ref{CBBdagger}). One could, of course, write the coefficients in terms of the bare coupling constant $q_0$, leading to hatted coefficient-functions. But we choose to present the period matrix in terms of the effective $q$ parameter since the results take a simple form. In \eqref{Omegaspapp}, the functions  $h_n(q,T_\ell)$ and $g_n(q,T_\ell)$ are the same ones obtained above while the $k_n(q,T_\ell)$ are given by
\begin{subequations}
\begin{align}
k_0 &= \frac{T_2^2}{2} \left(q-6 q^2+57 q^3-452 q^4+\ldots\right) \notag\\
&~~~-\frac{T_4}{12} \left(1+12 q+36 q^2+12 q^3+84 q^4+\ldots\right)~, \\
k_1 &= -\frac{T_2^2 T_3}{144} \left(1-162 q^2-2400 q^3+1512 q^4\right) \notag\\
&~~~+\frac{T_3 T_4}{72} \left(1-162 q^2-1248 q^3-4536 q^4\right) \notag  \\
&~~~+\frac{T_2 T_5}{60} \left(1-1788 q^3+2916 q^4+\ldots\right) ~, \\
k_2 &= -\frac{T_2^5}{8640} \left(1-540 q^2-5100 q^3-79380 q^4+\ldots\right) \notag\\
&~~~+ \frac{T_2^3 T_4}{864} \left(1-432 q^2-8592 q^3-39096 q^4+\ldots\right)\notag  \\
&~~~-\frac{T_3^2 T_4}{648} \left(1-216 q^2-1752 q^3-1512 q^4\right) \notag\\
&~~~-\frac{T_5^2}{450} \left(1+378 q^2+6492 q^3+74196 q^4\right) \notag \\
&~~~+\frac{T_2T_4^2}{4} \left(3q^2+67 q^3+411 q^4+\ldots\right) \notag\\
&~~~-\frac{T_2T_3 T_5}{270} \left(1-378 q^2-5370 q^3-27756 q^4+\ldots\right) \notag\\
&~~~ +\frac{T_2^2T_3^2}{432} \left(1-324 q^2-3528 q^3-11376 q^4+\ldots\right)\notag\\ &~~~-\frac{T_2^2T_6}{432} \left(1-216 q^2-6720 q^3-39744 q^4\right) \notag \\
&~~~-\frac{T_4 T_6}{216} \left(1+216 q^2+6384 q^3+60912 q^4\right) ~. 
\end{align}
\label{knq}
\end{subequations} 

\section{Modular forms of $\Gamma_1(3)$}
\label{secn:modular}
In this appendix we collect some useful formulas on the modular and quasi-modular forms used in the main
text.

We first define $f_{\pm}(\tau)$ to be
\begin{equation}
f_{\pm}(\tau) = \left(\frac{\eta^3(\tau)}{\eta(3\tau)}\right)^3\pm \left(\frac{3\eta^3(3\tau)}{\eta(\tau)}\right)^3~,
\label{f+-}
\end{equation}
where the Dedekind $\eta$-function is
\begin{equation}
\eta(\tau) = q^{\frac{1}{24}}\prod_{n=1}^{\infty}(1-q^n)~.
\label{eta}
\end{equation}
According to \cite{Koblitz,Apostol}, the modular functions of $\Gamma_1(3)$ (the congruence subgroup of
$\mathrm{Sl}(2,\mathbb{Z})$ defined in (\ref{gamma13})) form a ring generated by 
the two basis elements: $\{f_1(\tau), f_{-}(\tau)\}$, where 
\begin{equation}
f_1(\tau) = \big(f_+(\tau)\big)^{\frac{1}{3}} ~.
\label{f1app}
\end{equation}
These functions have the following Fourier expansions for small $q=\rme^{2\pi \ii \tau}$:
\begin{equation}
\begin{aligned}
f_1 &= 1+ 6q+6q^3+6q^4 + \cdots~,\\
f_- &= 1-36q-54q^2-252q^3-468q^4 + \cdots~.
\end{aligned}
\label{f1-q}
\end{equation}
Note that $f_1$ is a modular form of  $\Gamma_1(3)$ with weight 1, while $f_-$ is a modular form of
$\Gamma_1(3)$ with weight 3. This means is that under the $S'$-transformation
\begin{equation}
S'~~:~\tau~\rightarrow~\frac{\tau}{1-3\tau}~,
\label{S'app}
\end{equation}
these functions behave as follows
\begin{equation}
\begin{aligned}
f_1\left(\frac{\tau}{1-3\tau}\right) &= (1-3\tau)\, f_1(\tau)~,\\
f_-\left(\frac{\tau}{1-3\tau}\right) &= (1-3\tau)^3\, f_-(\tau)~.
\end{aligned}
\end{equation}
These modular forms have nice transformation properties also under the $S$-transformation
\begin{equation}
S~~:~~\tau ~\rightarrow~ -\frac{1}{3\tau}~,
\label{Sapp}
\end{equation}
which lies outside $\Gamma_1(3)$. To show this, we can exploit the known transformation 
of the Dedekind $\eta$-function
\begin{equation}
\eta\left(-\frac{1}{\tau}\right) = \sqrt{-\ii\tau}\, \eta(\tau)~,
\end{equation}
and check that the functions $f_{\pm}$ transform as
\begin{equation}
f_{\pm}\left(-\frac{1}{3\tau}\right) =\mp  (\ii\sqrt{3}\tau)^{3} \, f_{\pm}(\tau) ~.
\label{Sfpmapp}
\end{equation}
{From} this we get the $S$-transformation of $f_1$ which clearly is only determined up to a $\mathbb{Z}_3$ phase factor that we choose it to be 1. Thus,
\begin{equation}
f_1\left(-\frac{1}{3\tau}\right) =  -\, (\ii \sqrt{3}\tau)\, f_1(\tau)~.
\label{Sf1app}
\end{equation}

\subsubsection*{Basis of quasi-modular forms}
We now list the basis of quasi-modular forms of $\Gamma_1(3)$ which have been used in the main text. In addition to the weight under the $\Gamma_1(3)$ transformations, we have a $\mathbb{Z}_2$ 
charge, which we have call $S$-parity, determined by the properties under the $S$-transformation. 
For instance, from the transformation properties (\ref{Sf1app}) and (\ref{Sfpmapp})
we assign $f_1$ an $S$-parity $(-1)$ and $f_-$ an $S$-parity $(+1)$. 
In addition to these modular forms, we also have the quasi-modular form $E_2(\tau)$. However, $E_2$ by itself, does not 
transform well under the $S$-transformation; indeed
\begin{equation}
E_2 \left(-\frac{1}{3\tau}\right) = (3\tau)^2\, E_2(3\tau) + \frac{6}{\ii \pi}(3\tau) ~. 
\end{equation}
On the other hand one can prove the following triplication formula
\begin{equation}
E_2(3\tau) = \frac{1}{3}E_2(\tau) + \frac{2}{3}f_1^2(\tau)
\end{equation}
which shows that $E_2$ mixes with $f_1^2$ under the $S$-transformation. Using this fact and 
the $S$-transformation property of $f_1$ given in (\ref{Sf1app}), 
it is possible to check that the following combination 
\begin{equation}
\widetilde{E}_2(\tau) = E_2(\tau) + f_1^2(\tau)
\end{equation}
transforms into itself, save for an anomalous term:
\begin{equation}
\widetilde{E}_2\left(-\frac{1}{3\tau}\right) = - (\ii\sqrt{3}\tau)^2 \widetilde{E}_2(\tau) + \frac{18}{\ii \pi}\tau ~.
\end{equation}
Thus, we can assign the quasi-modular form $\widetilde{E}_2$ an $S$-parity equal to $-1$.

In Tab.~2 we list a basis of the modular forms of $\Gamma_1(3)$ up to weight 10, distinguishing them according to 
their $S$-parity.
\begin{table}[ht]
\begin{center}
\begin{tabular}{|c|c|c|}
\hline
Weight & Modular forms of $\Gamma_1(3)$         &  $S$-parity \\
\hline
1 & $f_1$ &$-1$\\
\hline
 2 &
$f_1^2$ &$+1$\\
\hline
3& $f_1^3$ & $-1$\\
&$f_-$
& $+1$\\
\hline
4 &  $f_1f_-$  & $-1$ \\
& $f_1^4$ \, & $+1$\\
\hline
5 & $f_1^5$  & $-1$\\
 & $f_1^2f_-$  & $+1$\\
\hline
6 & $f_1^3f_-$  & $-1$\\
 & $f_1^6$\,, \, $f_-^2$\,, \, & $+1$\\
\hline
7& $f_1^7$\,, \, $f_1f_-^2$ \, &$ -1$\\
& $f_1^4f_-$\,  & $+1$\\
\hline
8&$f_1^5f_-$\,  &$-1$\\
&$f_1^8$\,, \, $f_1^2f_-^2$ \, & $+1$\\
\hline
9 & $ f_1^9$\,, \, $f_1^3f_-^2$ & $-1$\\
& $f_1^6f_-$\,, \, $f_-^3$ &$+1$ \\
\hline
10& $f_1^7f_-$\,, \,  $f_1f_-^3$ &$-1$\\
&$f_1^{10}$\,, \, $f_1^4f_-^2$   & $+1$\\
\hline
\end{tabular}
\end{center}
\caption{A basis of modular forms of $\Gamma_1(3)$ up to weight 10, classified according to their $S$-parity.\label{Table2}}
\end{table}

\section{The coefficients $g_3$ and $g_4$}
\label{secn:appgn}

Here we report the expression of the coefficients $g_3$ and $g_4$ in the expansion of the
dual period, written in terms of the quasi-modular functions of $\Gamma_1(3)$, which we obtained by
solving the recursion relation (\ref{recursiongn}). 
\begin{subequations}
\begin{align}
g_3 &= -\frac{T_3^4}{7128}\left(\frac{605 f_1^8 \widetilde{E}_2}{1296}-\frac{55 f_1^4 \widetilde{E}_2^3}{576}-\frac{323f_1^7 f_-}{648} +\frac{55f_1^3 \widetilde{E}_2^2 f_-}{96} -\frac{275f_1^2 \widetilde{E}_2 f_-^2}{324} +\frac{119f_1 f_-^3}{162} \right) \notag\\
& ~~~ +\frac{T_3^2 T_6}{1188}\left(\frac{11f_1^8 \widetilde{E}_2}{32} -\frac{73 f_1^7 f_-}{144}+\frac{11f_1^3 \widetilde{E}_2^2 f_-}{32} -\frac{55f_1^2 \widetilde{E}_2 f_-^2}{96} +\frac{85f_1 f_-^3}{144} \right) \notag\\
& ~~~ +\frac{T_6^2}{792} \left(\frac{7f_1^7 f_-}{10} +\frac{11f_1^2 \widetilde{E}_2 f_-^2}{30} -\frac{13f_1 f_-^3}{30} \right) ~,\\
\notag\\
g_4 &= -\frac{T_3^5}{68040}\left(\frac{5435 f_1^{13}}{21384}-\frac{5005 f_1^9 \widetilde{E}_2^2}{10368}+\frac{5005 f_1^5 \widetilde{E}_2^4}{82944}+\frac{16835 f_1^8 \widetilde{E}_2 f_-}{14256}-\frac{5005 f_1^4 \widetilde{E}_2^3 f_-}{10368} \right. \notag\\ 
& \qquad \qquad \qquad \left. -\frac{91351 f_1^7 f_-^2}{85536}+\frac{5915 f_1^3 \widetilde{E}_2^2 f_-^2}{5184}-\frac{15925 f_1^2 \widetilde{E}_2 f_-^3}{9504}+\frac{182455 f_1 f_-^4}{171072}\right) \notag\\
&~~~ +\frac{T_3^3 T_6}{108}\left(\frac{13 f_1^{13}}{3960}-\frac{13 f_1^9 \widetilde{E}_2^2}{3840}+\frac{10621 f_1^8 \widetilde{E}_2 f_-}{855360}-\frac{143 f_1^4 \widetilde{E}_2^3 f_-}{51840}-\frac{7631 f_1^7 f_-^2}{855360} \right. \notag \\
& \qquad \qquad \qquad \left. +\frac{299 f_1^3 \widetilde{E}_2^2 f_-^2}{34560}-\frac{11791 f_1^2 \widetilde{E}_2 f_-^3}{855360}+\frac{8021 f_1 f_-^4}{855360}\right) \notag\\
& ~~~ -\frac{T_3 T_6^2}{1512}\left(\frac{237 f_1^{13}}{2200}+\frac{9373 f_1^8 \widetilde{E}_2 f_-}{26400}-\frac{5161 f_1^7 f_-^2}{39600}+\frac{91}{600} f_1^3 \widetilde{E}_2^2 f_-^2 \right. \notag \\
& \qquad \qquad \qquad \left. -\frac{2093 f_1^2 \widetilde{E}_2 f_-^3}{8800}+\frac{7189 f_1 f_-^4}{39600}\right) ~.
\end{align}
\end{subequations}
As is evident from the above expressions and in the earlier discussion on recursion relations, the quasi-modular piece of these expressions -- terms with $\tilde{E}_2$ dependence -- are completely fixed by the recursion relation. 
The remaining pieces are decided by considerations of weight and $S$-parity. Thus, to fix $g_3$ one needs to use the perturbative and  $1$-instanton contributions, while to fix $g_4$, one needs to take into account $2$-instanton corrections as well.

By expanding the quasi-modular forms in powers of $q$ one can check that the expressions agree with the instanton corrections up to $k=4$ given in (\ref{g3q}) and (\ref{g4q}).

\providecommand{\href}[2]{#2}\begingroup\raggedright\endgroup

\end{document}